\documentclass[aps,prl,reprint,amsmath,amssymb]{revtex4-2} 
\usepackage[utf8]{inputenc}
\usepackage{hyperref,graphicx,xcolor}

\graphicspath{{./}{./figures/}}

\def\L{ {\mathrm{L}} }

\def\O{ {\mathrm{O}} }

\def\R{ {\mathrm{R}} }
\def\d{ {\mathrm{d}} }
\def\e{ {\mathrm{e}} }

\def\i{ {\mathrm{i}} }

\def\vk{ {\boldsymbol{k}} }

\newcommand{\beq}{\begin{equation}}
\newcommand{\eeq}{\end{equation}}

\newcommand{\abs}[1]{\left| #1 \right|}

\newcommand{\diag}[1]{\text{diag}\left\{ #1 \right\}}

\newcommand{\wt}[1]{\widetilde{#1}}

\renewcommand{\O}{\mathrm{O}}

\newcommand{\ve}{\varepsilon}

\newcommand{\hlt}{\mathcal{H}}

\newcommand\prlsection[1]{\emph{#1.}---}
\newcommand\area{\mathcal{A}}

\begin{document}

\title{Quantum Oscillation Signatures of Fermi Arcs in Tunnel Magnetoconductance}

\author{Adam Yanis Chaou}
\author{Vatsal Dwivedi}
\author{Maxim Breitkreiz}
\email{breitkr@physik.fu-berlin.de}
\affiliation{Dahlem Center for Complex Quantum Systems and Fachbereich Physik, Freie Universit\"at Berlin, 14195 Berlin, Germany}

\begin{abstract}
Fermi-arc surface states of Weyl semimetals exhibit a unique combination of localization to a surface and connectivity to the bulk Weyl fermions that can move along the localization direction. We predict anomalous quantum-oscillation signatures of Fermi arcs in the tunnel mangetoconductance across an interface between two Weyl semimetals. These oscillations stem from a momentum-space analog of Aharonov-Bohm interference of electrons moving along the interface Fermi arcs, driven by an external magnetic field normal to the interface. The Fermi arcs' connectivity to the bulk enables their characterization via transport normal to the interface, while their localization manifests in a strong field-angle anisotropy of the oscillations. This combination distinguishes these anomalous oscillations from conventional Shubnikov-de Haas oscillations and makes them identifiable even in complex oscillation spectra of real materials. 
\end{abstract}

\maketitle

\prlsection{Introduction}
Weyl semimetals (WSMs) are a class of three-dimensional topological semimetals that host pairs of topologically protected gapless points that can be described as Weyl fermions at low energies\cite{Wan2011,Burkov2011,Xu2011,Xu2015, Xu2015b, Lv2015,Armitage2017, Yan2017, Burkov2017,Hasan2021, Bernevig2022a}. A remarkable feature of Weyl fermions is the chiral anomaly \cite{Adler1969,Bell1969}, which can be understood as a spectral flow along the chiral zeroth Landau level dispersing parallel to an applied magnetic field \cite{Nielsen1983}. The boundary manifestation of the bulk topology of WSMs are Fermi arcs --- lines of zero-energy surface states that connect projections of opposite-chirality Weyl nodes within the surface Brillouin zone\cite{Wan2011}. 

An interface between two WSMs also features Fermi arcs unless Weyl nodes of the same chirality from different WMSs project on top of each other \cite{Dwivedi2018,Abdulla2021,Mathur2022,Kaushik2022, Murthy2020, Kundu2023,Chaou2023}. Interface Fermi arcs either connect nodes of opposite chirality from the same WSM (as in the case of surface Fermi arcs), which we term \emph{heterochiral} connectivity, or nodes of identical chirality from different WSMs, which we call \emph{homochiral} connectivity. In the presence of a magnetic field normal to the interface (the ``longitudinal'' direction), anomalous charge current carried by the chiral Landau levels in longitudinal direction are redirected along the Fermi arcs by the Lorentz force. For homochiral Fermi arcs, this leads to perfect transmission of the anomalous charge current\cite{Chaou2023}, while for heterochiral Fermi arcs, it leads to perfect reflection and hence the vanishing of the tunnel conductance. In both cases, the Fermi arcs bear a unique combination of local and nonlocal qualities in that they are localized to the interface but mediate transport normal to it i.e., along their localization direction \cite{Parameswaran2014,Baum2015,Chaou2023}. 

Quantum-oscillations, such as the Shubnikov-de Haas (SdH) or de Haas-van Alphen effects, constitute standard experimental tools for mapping low-energy states of metals \cite{Shoenberg1984}. Identification of Fermi arcs using these  well-established techniques has, however, been challenging. In principle, Fermi arcs are detectable via quantum oscillations stemming from the so-called Weyl orbit \cite{Potter2014,Moll2016} ---  the cyclotron orbit of a thin WSM slab that involves the coherent motion along Fermi arcs on both surfaces of the slab, connected by chiral Landau levels across the slab width. However, in this case the characteristic nonlocality manifests itself only in the slab-width dependence of the oscillation shift, which, alongside the requirement of a small slab width to ensure phase coherence, makes the experimental identification of Fermi arcs this way very difficult \cite{Galletti2019,Kar2023}. Furthermore, typical WSM materials exhibit additional Fermi pockets, whose trivial orbits also contribute to the full quantum oscillation spectrum. 

\begin{figure}
    \includegraphics[width=0.85\columnwidth]{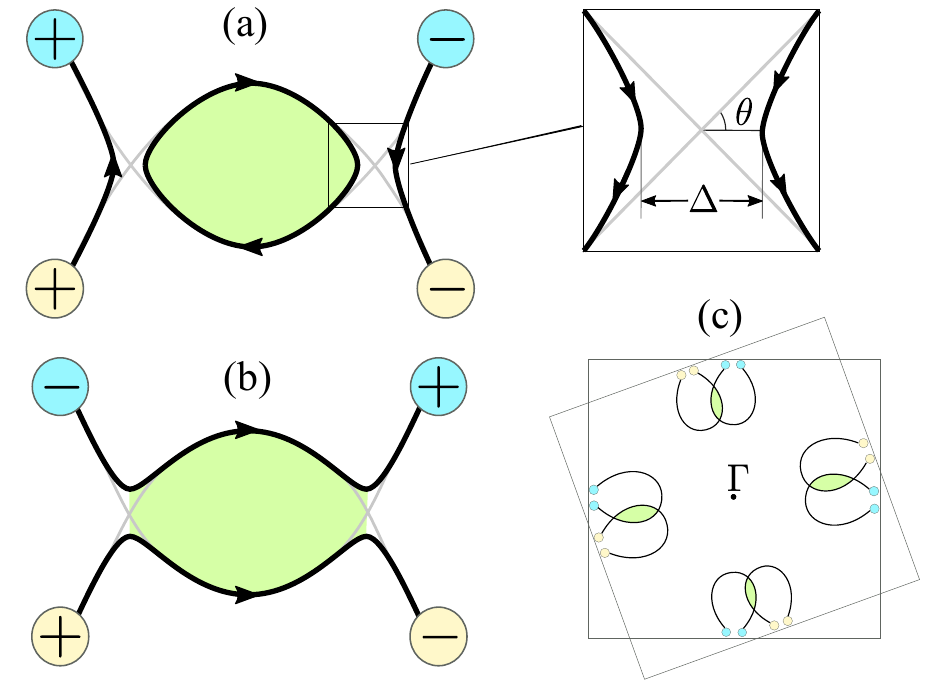}	
    \caption{Fermi-arc configurations (black lines) with (a) homochiral and (b) heterochiral connectivity, exhibiting two close encounters at a weakly-coupled interface between two WSMs. The projections of Weyl nodes from the two WSMs are depicted as yellow/blue circles (chirality indicated as $+/-$), the arrows indicate the direction of motion in a magnetic field out of the plane, and the (green) shaded region between the two close encounters determines the frequency of quantum oscillations. The inset shows a single close encounter with the minimum separation $\Delta$ and the opening angle $2\theta$. (c) Long crescent Fermi arcs of two TaAs (001) surfaces from \emph{ab initio} calculations \cite{Chang2016} rotated by $20$° with respect to each other. 
     }
    \label{fig:int_arcs}    
\end{figure}

In this article, we predict a characteristic quantum-oscillation signature of Fermi arcs in the tunnel magnetoconductance. We consider interfaces between two WSMs where Fermi arcs exhibit two or more close encounters, as exemplified in Fig.\ \ref{fig:int_arcs}. Such a Fermi-arc configuration can be experimentally realized, e.g., at an interface between two weakly coupled WSMs with curved Fermi arcs, which can be two different WSM materials or same materials rotated with respect to each other, such as rotated TaAs (001) surfaces \cite{Xu2015a,Chang2016} shown in Fig.\ \ref{fig:int_arcs} (c). We predict oscillations in the magnetoconductance as a function of the inverse longitudinal magnetic-field component, whose frequency depends on the momentum-space areas enclosed by the Fermi arcs. 
Due to magnetic breakdown at the  encounters \cite{Cohen1961,Blount1962,Kaganov1983}, there are multiple effective paths \cite{Falicov1966,VanDelft2018,Muller2020,Muller2022} connecting interface projections of the Weyl nodes. The field-induced motion along the Fermi arcs thus becomes subject to Aharonov-Bohm-like interference in momentum space. The resulting quantum oscillations of the tunnel magnetoconductance are fundamentally different from the possibly coexisting SdH oscillations of trivial states, stemming from Landau-quantized levels passing the Fermi energy. Experimentally they are identifiable by their characteristic field-angle anisotropy, which is a direct consequence of the unique local/nonlocal Fermi-arc character, and the spectrum of higher harmonics. In the following, we describe the proposed setup, compute the tunnel magnetoconductance semiclassically, and compare the predictions to exact numerical simulations on a lattice model. 

\prlsection{Fermi-arc mediated tunnel magnetoconductance}
We consider the tunnel magnetoconductance across an interface between two WSMs, for which at least some of the projections of Weyl nodes onto the interface Brillouin zone do not overlap, such that the interface features Fermi arcs connecting well-separated Weyl-node projections. More precisely, we require this separation to be larger than the inverse magnetic length $\ell_B^{-1}$ ($\sim 0.02\text{\AA}^{-1}$ for the maximum realistic magnetic field $B=30$T). The interface Fermi arcs result from the hybridization of the surface Fermi arcs of the two WSMs, so that their exact form depends on the specific Fermi-arc arrangements of the two coupled surfaces and the coupling strength. If the surface Fermi arcs from the two sides intersect, then a weak coupling generally leads to an avoided crossing in the momentum space (see Fig.\ \ref{fig:int_arcs} (a) and (b)), which we term a ``close encounter''.

A pair of interface Fermi arcs contributes tunnel conductance \cite{Chaou2023}
\begin{equation}
    G = \frac{e^2}{h} N_B T, 	
    \label{eq:G}
\end{equation}
where $e^2/h$ is the quantum of conductance,  $N_B\propto B$ is the Landau level degeneracy (number of flux quanta through the interface), and $0<T<1$ is the total transmission probability along the Fermi arcs. For a pair of homochiral (heterochiral) Fermi arcs that are well separated everywhere, (i.e., for separations $\gg \ell_B^{-1}$), the transmission probability is $T=1$ ($T=0$). In the former case, this implies a universal conductance independent of band details such as Fermi-arc shape, Fermi velocity, and Fermi energy.  However, if two Fermi arcs approach within $\sim\ell_B^{-1}$, then magnetic breakdown leads to suppression (enhancement) of the transmission probability for the homochiral (heterochiral) Fermi arcs. In particular, for large fields, this results in a transmission probability proportional to $1/B$, leading to the saturation of the conductance \cite{Chaou2023}. Unlike the contributions of trivial states to the magnetoconductance, the Fermi-arc contribution does not show SdH quantum oscillations because the Fermi-arc mediated current is carried exclusively by the lowest Landau levels of Weyl Fermions at all field strengths \cite{Breitkreiz2022}.

We now show that anomalous, \emph{non-SdH} quantum oscillations  occur for Fermi-arc arrangements with more than one close encounter. In case of two encounters, the two possible interface Fermi arc configurations are depicted in Fig.~\ref{fig:int_arcs} (a) and (b), which exhibt homochiral and heterochiral connectivity, respectively. In the absence of magnetic breakdown, these two configurations would yield $G=(e^2/h)N_B$ and $G=0$, respectively (in analogy with the single-encounter case \cite{Chaou2023}). We next compute the transmission probability in presence of magnetic breakdown. For clarity, we focus on the two-encounter case, but our analysis can be straightforwardly extended to more nodes and/or close encounters. 

\prlsection{Semiclassical analysis}
To compute the transmission amplitudes, we employ a semiclassical approach away from the close encounters coupled with the full quantum problem near them. Semiclassically, the electron wavepackets incident on the interface are driven along the Fermi arc by the Lorentz force until they encounter another Weyl node of identical/opposite chirality, leading to transmission/reflection across the interface. The quantum effects are encoded in the path-dependent Aharonov-Bohm, de Broglie, and Maslov phases picked up by them \cite{Keller1958}. Near a close encounter, the description of magnetic breakdown maps onto the Landau-Zener problem \cite{Landau1977,Kaganov1983}. Thus, the splitting of electron trajectories is 
described by the S-matrix \cite{Kaganov1983} 
\begin{equation}
    S(B) = 
    \begin{pmatrix}
        \sqrt{1-\e^{-\gamma}} \e^{\i\alpha} & -\i \sqrt{\e^{-\gamma}} \\ 
        -\i \sqrt{\e^{-\gamma}}& \sqrt{1-\e^{-\gamma}} \e^{-\i\alpha} 
    \end{pmatrix},
    \label{eq:smat_enc}
\end{equation}
where $\gamma = B_0/B$, $\e^{-\gamma}$ is the tunneling probability and 
\begin{equation}
    \alpha = \frac\pi4 + \frac\gamma{2\pi}\left[ 1 - \ln \left(\frac\gamma{2\pi} \right)\right] + \arg \Gamma \left(\frac{\i\gamma}{2\pi} \right) 
    \label{eq:alpha_def}    
\end{equation}
is the additional phase acquired by a state when it does \emph{not} tunnel. The scattering process is governed by a single free parameter, the breakdown field $B_0$, which is determined by the geometry of the close encounter as 
\begin{equation}
	B_0 = \frac\pi4 \Delta^2 \tan\theta,  
	\label{eq:B_0_def}
\end{equation}
where $\theta$ is the angle of intersection between the two Fermi arcs in the decoupled limit and $\Delta$ is the minimum separation (see inset of Fig.~\ref{fig:int_arcs} (a)). The total transmission amplitude is the sum over all paths weighted with the scattering amplitudes of the encounters given in \eqref{eq:smat_enc} and phase factors stemming from motion along the connecting Fermi-arc segments. 

For the heterochiral Fermi-arc configuration, the sum over two possible paths (see Supplemental Material (SM) for a detailed derivation) leads to the transmission probability 
\begin{equation}
    T_\text{het} = 2 \e^{-\gamma}(1-\e^{-\gamma}) (1 + \cos\phi),
    \label{eq:P_het}
\end{equation}
where 
\begin{equation}
	\phi = 2\alpha - \beta, \qquad 
	\beta = \frac{\area}{B} + \pi. 
	\label{eq:phi_def2}
\end{equation}
Here, $\beta$ is the difference between the de Broglie phases acquired along the two paths between the close encounters, given by the momentum space area $\area$ enclosed by the Fermi arcs between the two points of minimum separation at the Fermi level. The additional $\pi$ in $\beta$ is a Maslov phase \cite{Keller1958} corresponding to the two classical turning points encountered in going around the loop. The magnetoconductance oscillates as a function of $B^{-1}$, with a slowly varying envelope given by $T_\text{het}^\text{max} =  4 \e^{-\gamma}(1-\e^{-\gamma})$. For $B\to0$, we get an exponentially suppressed $T_\text{het} \simeq 4 e^{-\gamma}$, vanishing as expected for interface Fermi arcs connecting the nodes from the same WSM. For $B\to\infty$, $T_\text{het}^\text{max} \approx 4B_0/B$, so that the conductance saturates at a value proportional to $B_0$. 

For the homochiral configuration, there are infinitely many paths leading to transmission, corresponding to tunneling into the loop of zero modes, traversing it arbitrarily many times, and finally tunneling out of it on the same side of the loop. Summing over these possibilities, we obtain the transmission probability as 
\begin{equation}
    T_\text{hom} = 1-\frac{\e^{-2\gamma}}{\e^{-2\gamma} + 2(1-\e^{-\gamma}) (1-\cos\phi)}, 
    \label{eq:P_hom}
\end{equation}
where 
\begin{equation}
    \phi = 2\alpha + \beta, \qquad 
    \beta = \frac{\area}{B} + \pi. 
    \label{eq:phi_def}
\end{equation}
Here, $\beta$ is the total semiclassical phase acquired on traversing the loop of zero modes (in opposite direction compared to the phase difference of the heterochiral case), whereby the first term comes from the sum of Aharanov-Bohm and de Broglie phases given by the momentum space area $\area$ enclosed by the loop, and the additional Maslov phase $\pi$. The magnetoconductance also oscillates as a function of $B^{-1}$, with the envelope given by $T_\text{hom}^\text{max} = 4(1 - \e^{-\gamma})/(2- \e^{-\gamma})^2$. For $B\to0$, we get $T_\text{hom} = 1$ as expected for homochiral Fermi arcs, leading to a linear-in-$B$ magnetocondutance. The limit $B\to\infty$ again leads to saturation of the magnetoconductance  $T_\text{hom}^\text{max} \approx 4B_0/B$. 

\begin{figure}
    \includegraphics[width=\columnwidth]{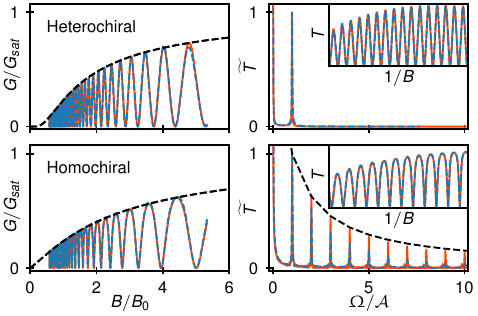}   
    \caption{Left column: Conductance as a function of magnetic field for heterochiral (top row) and homochiral (bottom row) connectivity obtained from analytics (solid orange lines) and numerics (dashed blue lines). Right: Fourier transform of tunnel probability $\widetilde{T}(\Omega)$ normalized to $\widetilde{T}(\area)=1$  (inset shows $T(1/B)$). The dashed black line shows the analytically obtained damping of higher harmonics. Model parameters are $\kappa = 0.07$ and $\ve_F = 0.1$ (others in text).
    } 
    \label{fig:G_vs_B}
\end{figure}

In Fig.\ \ref{fig:G_vs_B}, we plot the conductance (given by Eq.~\ref{eq:G}) as well as the tunnel probability and its Fourier transform for a specific value of parameters and compare them with numerical simulations on a lattice model (detailed below). While the conductance shows qualitatively similar features for both homochiral and heterochiral connectivities, they can be easily distinguished by the Fourier transform. For the heterochiral connectivity, as the oscillations result from the interference of only two paths, the Fourier transform exhibit a single peak, with the frequency corresponding to the area $\area$ enclosed between the two Fermi arcs. On the other hand, for homochiral connectivities, the trivial loop connecting the homochiral Fermi arcs can be traversed multiple times (similar to SdH oscillations), leading to harmonics at frequencies $n\area$. In contrast to SdH oscillations, however, for each traversal of the loop, there is a nonzero probability of tunneling out of the loop to the opposite Fermi arc, leading to reflection from the interface. Thus, the the higher harmonics are damped, with the height of the $n^\text{th}$ peak proportional to $\left[ \Psi \left( n/2+1  \right) - \Psi \left( n/2+1/2 \right)  \right]$, where $\Psi(z) = \Gamma'(z)/\Gamma(z)$ is the digamma function (see SM for the derivation). This damping profile exhibits a long tail and should be visible even for small $\area$. 

\prlsection{Lattice model and numerical results} 
We compare our analytical predictions with a numerical simulation on an explicit lattice model, for which we compute the model-specific parameters $\Delta$, $\theta$, and $\area$ that enter the analytical formula, so that analytics and numerics can be compared without any fitting parameters.
We consider a Bloch Hamiltonian of the form \cite{Dwivedi2016b}
\begin{align} 
	\hlt(\vk) = \hlt_x(k_x) +  \eta_y(\vk_\perp) \tau^y + \eta_z(\vk_\perp) \tau^z,
	\label{eq:hlt_orig} 
\end{align} 
where the Pauli matrices $\tau^a$ represent a pseudospin degree of freedom and 
\begin{equation}
	\hlt_x(k_x) = \sin k_x \tau^x + (1 - \cos k_x) \tau^z.
\end{equation}
The lattice constant and the hopping strength along $x$ are set to unity. This lattice model has Weyl nodes in the $k_x = 0$ plane at transverse momenta $\vk_\perp$ that satisfy $\eta_y(\vk_\perp) = \eta_z(\vk_\perp) =  0$. For a surface normal to $x$, the Fermi arcs are given by \cite{Dwivedi2016b} $\eta_y(\vk_\perp) = 0$, which exist only for lattice momenta satisfying $\eta_z(\vk_\perp) < 0$. 

We consider an interface between two WSMs that are described by the lattice model above with 
\begin{align}
	\eta_y^A (\vk_\perp) &= \xi_A \left( \cos k_y - \cos b_y + \zeta_A \sin k_z - \sin b_z \right), \nonumber \\ 
	\eta_z^A (\vk_\perp) &= \cos b_z - \cos k_z,   \label{eta_def}
\end{align}
where $A \in \{\L,\R\}$, $b_y, b_z \in (0,\pi)$ and $\xi_A, \zeta_A \in \{\pm1\}$. The two WSMs have Weyl nodes at $\vk_A = (0, \pm b_y, \zeta_A b_z)$ with chiralities $\chi = \pm \xi_A \zeta_A$. We model the tunnel junction by modulating the hopping along $x$ at the interface by a factor $0 \leq \kappa \leq 1$. To ensure that in the decoupled limit ($\kappa=0$), the Fermi arcs of the two sides intersect at two points, we set $\zeta_\L = -\zeta_\R = -1$. The intersection point is then given by $\vk_\perp = (\pm b_0, 0)$ with $b_0 \equiv \cos^{-1} \left(\cos b_y + \sin b_z\right)$. For $\kappa>0$, the Fermi arc connectivity is homochiral if $\xi_\L = -\xi_\R = 1$ and heterochiral if $\xi_\L = \xi_\R = 1$. We hereafter set $b_y = 3\pi/4$ and $b_z = \pi/2$. 

We obtain $\theta$ by linearizing the Fermi arc contours $\eta_y^{\L/\R}(\vk_\perp)=0$ about the intersection points $(\pm b_0, 0)$. This yields $q_z \approx -\zeta_{\L/\R} \sin b_0 q_y $, so that $\theta$ is given by  $\tan\theta=|q_z/q_y|=|\sin b_0|$ ($\tan\theta=|q_y/q_z|=|\csc b_0|$ ) for homochiral (heterochiral) connectivity. To compute $\Delta$ and $\area$, we employ the generalized transfer matrices \cite{Dwivedi2016, Dwivedi2016b}, as detailed in SM. This yields an implicit expression for the interface Fermi arcs in terms of $\ve$, $\vk_\perp$, and $\kappa$. Using the fact that the minimum separation $\Delta(\kappa)$ occurs along the lines $k_z = 0$ and $k_y = \pm b_0$ for the homochiral and heterochiral cases, respectively, we obtain 
\begin{align}
    \Delta_\text{hom}(\kappa) 
    &= \cos^{-1} (\cos b_0 - \kappa) - \cos^{-1} (\cos b_0 + \kappa) 
    \\     &\approx 2 \kappa \csc b_0 + \O(\kappa^3), \nonumber 
\end{align}
and 
\begin{align}
    \Delta_\text{het}(\kappa) 
    &= 2 \left[ \cos^{-1} \left( \frac{\kappa}{1 + \kappa^2} \right) - \tan^{-1} \left( \frac{1 - \kappa^2}{2\kappa} \right)  \right] 
\\   &\approx 2 \kappa + \O(\kappa^3). \nonumber 
\end{align}
The computation of $\area$ is analytically intractable, so that we obtain it by numerically integrating the implicit condition for the Fermi arcs at $\ve = \ve_F$. Inserting the expressions for $\Delta$ and $\tan\theta$ into Eq.~\eqref{eq:B_0_def}, we obtain the breakdown field which, together with $\area$, determines the analytic magnetoconducatance via Eqns.~\eqref{eq:P_het} and \eqref{eq:P_hom} inserted into \eqref{eq:G}. The exact numerical computation of the magnetoconductance is performed using the \emph{Kwant} package \cite{Groth2014}. We find perfect agreement between the analytics and numerics, as exemplified in Fig.~\ref{fig:G_vs_B}, for the relevant range of parameters $\kappa$, $\ve_F$, and $B$,  set by the requirement $\area\gg l_B^{-2}\sim l_{B_0}^{-2}$ to ensure observable oscillations (not too small oscillation frequencies) and uncoupled Weyl-node projections.

\prlsection{Discussion and conclusions}
\label{sec:conc}
We have demonstrated anomalous quantum oscillations in the tunnel magnetoconductance across a Weyl semimetal interface, arising from Aharonov-Bohm like interference effects enabled by magnetic breakdown in Fermi-arc networks. As the oscillations appear in the electron transport normal
to the interface and thus along the Fermi-arc localization direction, they manifest the unique
combination of local and nonlocal qualities of Fermi arcs.

The experimental fingerprint of the anomalous oscillations is an extreme field-angle anisotropy: The Fermi-arc contributions only depend on the longitudinal field component, which distinguishes them from other quantum oscillations that may arise from various two- or three-dimensional trivial Fermi pockets. The former lead to oscillations also in the transverse magnetic-field components, while the latter do not contribute to transport along their localization direction. 
This Fermi-arc signature appears to be better accessible than the width-dependence of SdH oscillation shifts of Weyl-orbits.

Further peculiarities lie in the behavior of higher harmonics of the anomalous oscillations, which, moreover, allow to distinguish different Fermi-arc connectivities. For heterochiral connectivity, the spectrum does not feature higher harmonics (unlike SdH oscillations), while for homochiral connectivity, the higher harmonics feature unusual damping stemming from magnetic breakdown.

As the Fermi arc contribution to quantum oscillations requires coherent transport along entire loops of the interface Fermi arcs, we expect them to be sensitive to temperature- and disorder-induced decoherence, similar to conventional SdH oscillations, which sets a lower bound on the magnetic field for observation of the oscillations \cite{Shoenberg1984}. However, this is less restrictive to the constraints for the observation of Weyl orbits, which require coherent motion across both the bulk and the surface of a WSM thin film \cite{Potter2014}. 

\begin{acknowledgments}
\prlsection{Acknowledgments}
We thank P.W.\ Brouwer, L.I.\ Glazman, M.\ Mansouri, S.A.\ Parameswaran, and S.\ Wiedmann,  for useful discussions. This research was funded by the Deutsche Forschungsgemeinschaft (DFG, German Research Foundation) through CRC-TR 183 “Entangled States of Matter” and the Emmy Noether program, Project No. 506208038.
\end{acknowledgments}

\pagebreak
\begin{widetext}
\section*{Supplemental Material}      
\appendix 

\setcounter{equation}{0}
\renewcommand{\theequation}{S\arabic{equation}}
\setcounter{figure}{0}
\renewcommand{\thefigure}{S\arabic{figure}}

\section{Computation of the S-matrix}
\label{app:smat}
In this section, we compute the total S-matrix for the interface. More explicitly, we compute $S_\text{full}$ defined as 
\begin{equation}
	\begin{pmatrix} f_1 \\ f_2 \end{pmatrix}
	= S_\text{full}
	\begin{pmatrix} i_1 \\ i_2 \end{pmatrix}, 
\end{equation}
where $i_{1,2}$ and $f_{1,2}$ are defined in Fig.~\ref{fig:smat}. To this end, we use the S-matrix for a single encounter given by Eq.~\eqref{eq:smat_enc}, as well as the semiclassical phase relating the internal (dashed) inputs. 

\paragraph{Heterochiral connectivity.---}
In this case, the wavefunctions on the internal legs are related as $i_a' = \e^{\i\beta_a} f_a'$ for $a = 1,2$. The full S-matrix is thus given by 
\begin{align}
    S_\text{full} 
    = S_0 
    \begin{pmatrix}
        \e^{\i\beta_2} & 0 \\ 
        0 & \e^{\i\beta_1}
    \end{pmatrix}
    S_0 
    = \e^{\i\beta_0} 
    \begin{pmatrix}
        \e^{\i\alpha} \left( \e^{\i\phi/2} - 2 W \cos\frac\phi2 \right) &-2\i \sqrt{W(1-W)} \cos\frac\phi2  \\ 
        -2\i \sqrt{W(1-W)} \cos\frac\phi2  & \e^{-\i\alpha} \left( \e^{-\i\phi/2} - 2W \cos\frac\phi2 \right)
    \end{pmatrix},
\end{align}
where $W = \e^{-B_0/B}$, $\beta_0 = \frac12(\beta_1 + \beta_2)$ and $\phi = 2\alpha - \beta$, with $\beta \equiv \beta_2-\beta_1$. The amplitude for transmission across the interface is given by the \emph{off-diagonal} term of this S-matrix since the transmission across the interface is associated with transmission to a Weyl-node projection of the same chirality  (see Fig.~\ref{fig:smat}), so that 
\begin{equation}
	T = 4 W (1-W) \sin^2 \left( \frac\phi2 \right) = 2 W (1-W) (1 + \cos\phi). 
\end{equation}
The transmission probability exhibits an oscillatory behavior, with local maxima for $\phi = 2n\pi$ and zeros for $\phi = (2n+1)\pi$. The global maxima occurs near $W=1/2$, whereby $B = B_0/\ln 2$. 

\paragraph{Heterochiral connectivity.---}
In this case, we first compute the matrix $M$ such that 
\begin{equation}
    \begin{pmatrix} f_a \\ f_a' \end{pmatrix}
    = S
    \begin{pmatrix} i_a \\ i_a' \end{pmatrix}
    \iff 
    \begin{pmatrix} i_a' \\ f_a' \end{pmatrix}
    = M
    \begin{pmatrix} i_a \\ f_a \end{pmatrix}; 
    \qquad a = 1,2.
\end{equation} 
Explicitly, rearranging Eq.~\eqref{eq:smat_enc}, we get 
\begin{equation}
    M = -\frac\i{\sqrt{W}}
    \begin{pmatrix}
        \sqrt{1-W}\e^{-\i\alpha} & -1 \\ 
        1 & -\sqrt{1-W}\e^{\i\alpha}
    \end{pmatrix}. 
\end{equation}
The wavefunctions on the internal legs are related as $i_2' = \e^{\i\beta_1} f_1'$ and $f_2' = \e^{\i\beta_2} i_1'$. Thus, we can compute the matrix $M_\text{full}$ that relates $(i_2, f_2)^T$ to $(i_1, f_1)^T$ as 
\begin{equation}
    M_\text{full} = M^{-1} 
    \begin{pmatrix}
        0 & \e^{\i\beta_2} \\ 
        \e^{\i\beta_1} & 0
    \end{pmatrix} 
    M = -\frac{\e^{\i\beta_0}}W
    \begin{pmatrix}
        -2\i \sqrt{1-W} \sin\frac\phi2  & \e^{-\i \alpha} \left( W \e^{-\i\phi/2} + 2\i\sin\frac\phi2 \right) \\ 
        \e^{\i \alpha} \left( W \e^{-\i\phi/2} - 2\i\sin\frac\phi2 \right)  & 2\i \sqrt{1-W} \sin\frac\phi2 
    \end{pmatrix},
\end{equation}
where we now have $\phi = 2\alpha - \beta_1 + \beta_2 = 2\alpha + \beta$, with $\beta \equiv \beta_2-\beta_1$ being the total semiclassical phase acquired by traversing the loop \emph{counterclockwise}. The full S-matrix is thus given by 
\begin{equation}
    S_\text{full} 
    = -\frac{\e^{\i\alpha}}{W \e^{-\i\phi/2} + 2\i\sin\frac\phi2}
    \begin{pmatrix}
        -2\i \sqrt{1-W} \sin\frac\phi2  & W \e^{-\i \beta_0} \\ 
        W \e^{\i\beta_0}  & -2\i \sqrt{1-W} \sin\frac\phi2 
    \end{pmatrix}. 
    \label{eq:Sfull_hom}
\end{equation}
The amplitude for transmission across the interface is given by the \emph{diagonal} term of this S-matrix (see Fig.~\ref{fig:smat}), so that 
\begin{align}
    T &= \frac{4(1-W) \sin^2 \left( \frac\phi2 \right)}{W^2 + 4(1-W) \sin^2 \left( \frac\phi2 \right)}  
    = 1-\frac{W^2}{W^2 + 2(1-W) (1-\cos\phi)}.
\end{align}
Thus, the transmission probability exhibits an oscillating behavior with local maxima at $\phi = (2n+1)\pi$ and zeros at $\phi = 2n\pi$, in contrast to the heterochiral case. 

This transmission probability can also be computed by summing over the possible paths for transmission/reflection. It is somewhat easier to compute the amplitude for reflection. Using the amplitude for tunneling $-\i W$ and that for reflection $\sqrt{1-W} \e^{\i\alpha}$ at each close encounter, the reflection amplitude can be written as 
\begin{equation}
    r = -W \e^{\i\beta_1} \sum_{n=0}^\infty (1-W)^n \e^{-\i n\phi}, 
    \label{eq:ref_amp}
\end{equation}
where the $n^\text{th}$ term in the series corresponds to tunneling into the loop from the left side, going around it $n+\frac12$ times, and tunneling out from the right. Performing the infinite sum, we get 
\begin{equation}
    r = - \frac{W \e^{\i\beta_1}}{ 1 - (1-W) \e^{-\i \phi}}
      = - \frac{W \e^{\i(\beta_1+\phi/2)}}{ W \e^{-\i \phi/2} + 2\i\sin\frac\phi2},  
\end{equation}
which equals $[S_\text{full}]_{21}$ derived in Eq.~\eqref{eq:Sfull_hom}. 

\begin{figure}
    \includegraphics[width=0.65\columnwidth]{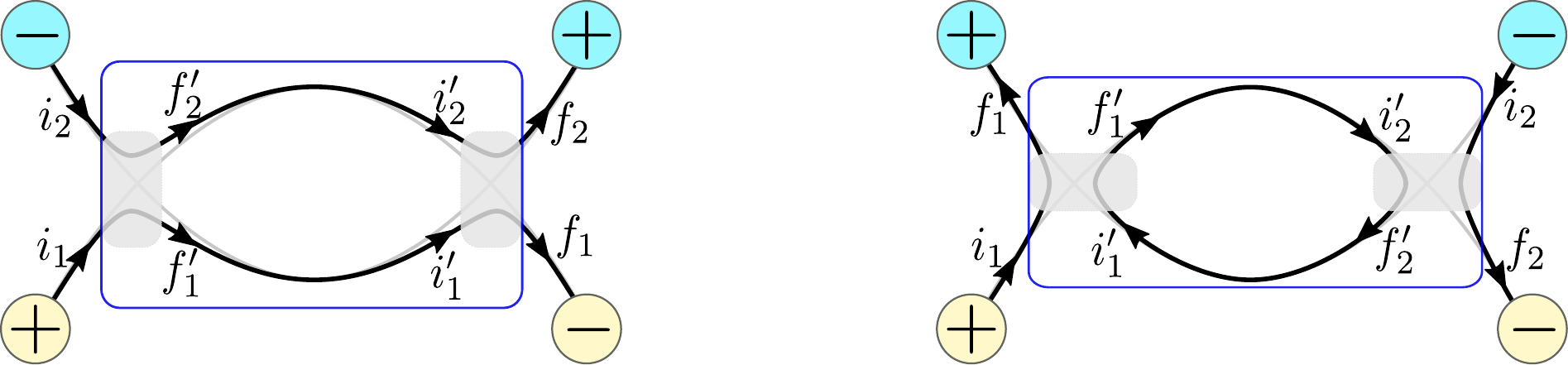}	
    \caption{The setup for the computation of the full S-matrix $S_\text{full}$ for the heterochiral (left) and homochiral (right) connectivities. In both cases, the gray rectangle represents the scattering region for single close encounter, whose S-matrix is given in Eq.~\eqref{eq:smat_enc}, while the blue rectangle represents the scattering regions for $S_\text{full}$. }
    \label{fig:smat}
\end{figure}

\subsection{Fourier transform of the conductance}
\label{app:fourier}
\newcommand\invB{u}  

We now compute the Fourier transform of the homochiral transmission probability as a function of $u \equiv 1/B$. As this probability is invariant under flipping the direction of the magnetic field, $T(\invB) = T(\abs{\invB})$, so that the Fourier transform is given by $\mathcal{F}[T] = \wt{T}(\Omega) + \wt{T}(-\Omega)$, where  
\begin{equation}
    \wt{T}(\Omega) \equiv \int_0^\infty \d\invB \, T(\invB) \e^{-\i\Omega\invB}.
\end{equation}
As $T(\invB) = 1 - R(\invB)$ and computing the Fourier transform of $R(\invB)$ is somewhat simpler, we first compute $\wt{R}(\Omega)$, in terms of which 
\begin{equation}
    \wt{T}(\Omega) = \pi \delta(\Omega) - \wt{R}(\Omega).
\end{equation}
To compute $\wt{R}(\Omega)$, we use Eq.~\eqref{eq:ref_amp} to write the reflection probability as a sum 
\begin{equation}
    R(\invB) = \abs{r}^2 = W^2 \sum_{m,n = 0}^\infty (1-W)^{m+n} \e^{\i(m-n)\phi}, 
\end{equation}
where $W = \e^{-B_0 \invB}$ and $\e^{\i\phi} = \e^{\i\area\invB}$, whereby we have ignored the $2\alpha$ contribution to $\phi$ since it is slowly varying as compared to $\area \invB$ and thus behaves as a constant phase shift when $\area \gg B_0$. Thus
\begin{align}
    R(\invB) 
    &= \sum_{m,n = 0}^\infty (1-\e^{B_0\invB})^{m+n} \e^{-2 B_0\invB + \i(m-n)\area \invB} 
    = \sum_{r=-\infty}^\infty \sum_{n=0}^\infty (1-\e^{B_0\invB})^{r+2n} \e^{-2 B_0\invB + \i r \area \invB}. 
\end{align}
We can now compute 
\begin{align}
    \wt{R}(\Omega) 
    &= \frac1{B_0} \sum_{r=-\infty}^\infty \sum_{n=0}^\infty \int_0^1 \d W W^{1 + \i (r\area - \Omega)/B_0} (1-W)^{r+2n} \nonumber \\ 
    &= \frac1{B_0} \sum_{r=-\infty}^\infty \sum_{n=0}^\infty \frac{(r+2n)! \Gamma(2 + \i \frac{r\area - \Omega}{B_0})}{\Gamma(3 + r + 2n + \i \frac{r\area - \Omega}{B_0}) }, 
\end{align}
where we identify the integral as the Euler beta function and use its relation to the gamma function. The $r=n=0$ term is 
\begin{equation}
    \frac{\Gamma(2 - \i \Omega/B_0)}{\Gamma(3 - \i \Omega/B_0) } = \frac1{2 - \i \Omega/B_0}.
\end{equation}
For $r \neq 0$, the magnitude of the individual terms is 
\begin{align}
    \abs{ \frac{(r+2n)! \Gamma(2 + \i \frac{r\area - \Omega}{B_0})}{\Gamma(3 + r + 2n + \i \frac{r\area - \Omega}{B_0}) } }^2 = \frac1{r + 2n + 1} \prod_{m=1}^{r+2n+1} \frac{m}{(m+1)^2 + \left( \frac{r\area - \Omega}{B_0} \right)^2}, 
\end{align}
where $x = (r\area - \Omega)/B_0$ and we have used $\abs{\Gamma(1+\i x)}^2 = \pi x / \sinh(\pi x)$ to write 
\begin{equation}
    \abs{\Gamma(n+\i x)}^2 = \frac{\pi x}{\sinh(\pi x)} \prod_{m=1}^{n-1} (m^2 + x^2). 
\end{equation}
Thus, $\wt{R}(\Omega)$ is sharply peaked when $\Omega = \nu\area$ for $\nu\in\mathbb{Z}$. The dominant contributions to the sum over $r$ thus come from $r=\nu$ as these terms are positive real numbers whereas terms $r\neq\nu$ are smaller in magnitude and have oscillating phases. The height of these peaks $\wt{R}_\nu \equiv \wt{R}(\nu\area)$ is given by 
\begin{align}
    \wt{R}_\nu 
    &= \frac1{B_0} \sum_{n=0}^\infty \frac{(\nu+2n)!}{(\nu+2n+2)!} = \frac1{B_0} \sum_{n=0}^\infty \frac1{(\nu+2n+1)(\nu+2n+2)},
\end{align}
where we have ignored the contribution from the $r=n=0$ term $\sim B_0/\area$ since $B_0 \ll \area$. The above sum can be evaluated by rewriting it as 
\begin{align}
    \wt{R}_\nu
    &= \frac1{2 B_0} \sum_{n=0}^\infty \left( \frac1{n + \frac{\nu+1}2} - \frac1{n + \frac{\nu}2 + 1}  \right)
    = \frac1{2 B_0} \left[ \Psi \left( \frac{\nu}2 + 1 \right) - \Psi \left( \frac{\nu+1}2 \right)  \right],
\end{align}
where $\Psi(z) = \Gamma'(z)/\Gamma(z)$ is the digamma function with the series representation 
\begin{equation}
    \Psi(z) = -\gamma + \sum_{n=1}^\infty \left( \frac1{n+1} - \frac1{n+z} \right), 
\end{equation}
with $\gamma$ being the Euler–Mascheroni constant. Using the recurrence relation $\Psi(z+1) = \Psi(z) + 1/z$, we can relate the strengths of the consecutive Fourier peaks as $\wt{R}_\nu = 1/(\nu B_0) - \wt{R}_{\nu-1}$. Finally, as $\wt{T}(\Omega)$ is identical to $\wt{R}(\Omega)$ for $\Omega \neq 0$, it also exhibits peaks of strength $\wt{T}_\nu = \wt{R}_\nu$ at $\Omega=\nu\area$.

\section{Transfer matrices for the model Hamiltonian}
\label{app:tmat}
\newcommand\tmat{\mathcal{T}}

In the main text, we consider Hamiltonians of the form 
\begin{align} 
	\hlt(\vk) =&\; \sin k_x \tau^x +  \eta_y(\vk_\perp) \tau^y + \left( 1 - \cos k_x + \eta_z(\vk_\perp) \right) \tau^z. 
	\label{bloch_hlt}
\end{align} 
The corresponding generalized transfer matrix along $x$ is given by \cite{Dwivedi2016, Dwivedi2016b}
\begin{equation} 
    \tmat = 
    \frac1{1+\eta_z}
    \begin{pmatrix}
        \ve^2 - \eta_y^2 - (1 + \eta_z)^2 \;\; &   -\ve+\eta_y \\ 
        \ve+\eta_y        & -1 
    \end{pmatrix}. 
    \label{TM}
\end{equation} 
For an interface along $x$ with tunneling strength $\kappa$, a necessary condition for the existence of a mode localized at the interface is \cite{Dwivedi2018}
\begin{equation} 
    \det \left[ \tmat_\L, K^{-1} \tmat_\R K \right] = 0,    
    \label{matching_cond}
\end{equation} 
where $\tmat_{\L,\R} $ denote the transfer matrices on the two sides of the interface and $K = \diag{(1/\kappa), \kappa }$. To simplify this further, we set 
\begin{equation}
    \tmat_A = 
    \begin{pmatrix}
        a_A & b_A \\ 
        c_A & d_A
    \end{pmatrix}, 
    \qquad \delta_A \equiv a_A - d_A; 
    \quad A \in \{\L,\R\}.
\end{equation}
Substituting in Eq.~\eqref{matching_cond}, after some algebra we arrive at 
\begin{equation}
    \beta^2 u^2 +  \beta \delta_\L \delta_\R u - \left( b_\L c_\L \delta_\R^2 + \delta_\L^2 b_\R c_\R + 4 \beta^2 \right) = 0,  
    \label{matching_cond2}
\end{equation}
where $u = \zeta + \zeta^{-1}$ and $\beta$ and $\zeta$ are chosen such that 
\begin{equation}
    \kappa^2 c_\L b_\R = \beta \zeta, \quad 
    \kappa^{-2} b_\L c_\R = \beta \zeta^{-1}.  
    \label{zeta_beta_def}
\end{equation}
To compute the area enclosed by the loop $\area$ for a fixed $\ve$ and $\kappa$, we numerically solve Eq.~\eqref{matching_cond2} to obtain $k_z(k_y)$ in the first quadrant, which we numerically integrate. The computation of the minimum separation $\Delta$ is, however, analytically tractable, as we now discuss. 

\subsection*{Computing the minimum arc separation}
From symmetry, we anticipate that the minimum separation between various branches of the interface Fermi arc lies along $k_z = 0$ for the homochiral case, whereby 
\begin{align}
    \eta_y^{\L/\R} (k_y,0) &= \pm (\cos k_y - \cos b_0), \nonumber \\  
    \eta_z^{\L/\R} (k_y,0) &= \cos b_z - 1.  
\end{align}
Similarly, for the heterochiral case, the minimum separation lies along $k_y = b_0$, whereby  
\begin{align}
    \eta_y^{\L/\R} (b_0, k_z) &= \pm \sin k_z, \nonumber \\  
    \eta_z^{\L/\R} (b_0, k_z) &= \cos b_z - \cos k_z.  
\end{align}
Thus, in both cases, $\eta_y^\L = -\eta_y^\R$ and $\eta_z^\L = \eta_z^\R$. Thus, for both cases, we can write the transfer matrices as 
\begin{equation}
    \tmat_\L = 
    \begin{pmatrix}
        a & b \\ 
        c & d
    \end{pmatrix}, 
    \quad 
    \tmat_\R = 
    \begin{pmatrix}
        a & -c \\ 
        -b & d
    \end{pmatrix}. 
\end{equation}
Eq.~\eqref{zeta_beta_def} now reduces to $\beta \zeta = -\kappa^2 c^2$, $\beta \zeta^{-1} = -\kappa^{-2} b^2$, so that we can choose 
\begin{equation}
    \zeta = \frac{c}{b} \kappa^2, \quad 
    \beta = -bc.
\end{equation}
Eq.~\eqref{matching_cond2} thus becomes 
\begin{equation}
    bc u^2 - \delta^2 u - 2 \left( \delta^2 + 2 bc \right) = 0, 
\end{equation}
in deriving which we have assumed that $b$ and $c$ are nonzero, since $b=0$ and $c=0$ yield the interface spectrum in the decoupled case, and we are interested in $\kappa\neq0$. We solve the quadratic in $u$ to get 
\begin{equation}
    u = 2\left( 1 + \frac{\delta^2}{2bc} \right), -2.
\end{equation}
But $u=-2 \implies \zeta=-1$, which yields $\eta_y=0$ for $\ve=0$. Since we expect the zero crossing of the interface spectrum to depend on $\kappa$, we ignore this branch. Thus, we are left with 
\begin{align}
    u = 2 + \frac{\delta^2}{bc} 
    \implies c^2 \kappa^2 + b^2 \kappa^{-2} - 2 bc = \delta^2 
    \implies c\kappa - b\kappa^{-1} = \mu\delta, 
\end{align}
where $\mu=\pm1$. Substituting 
\begin{equation} 
    \{b,c,\delta\} = \left\{ \frac{-\ve+\eta_y}{1+\eta_z}, \frac{\ve+\eta_y}{1+\eta_z}, \frac{\ve^2 - \eta_y^2 - \eta_z(2+\eta_z)}{1+\eta_z} \right\} , 
\end{equation}
we get 
\begin{equation}
    \ve^2 - 2\mu\ve\cosh\lambda - \eta_y^2 + 2\mu\eta_y\sinh\lambda - \eta_z(2+\eta_z) = 0, 
\end{equation}
where we have set $\kappa = \e^{-\lambda}$ with $\lambda\in[0,\infty)$. The interface spectrum is thus given by 
\begin{equation}
    \ve = \mu\cosh\lambda + \mu' \sqrt{(\eta_y - \mu\sinh\lambda)^2 + (1+\eta_z)^2}. 
\end{equation} 
We get four branches of solutions since $\mu, \mu' =\pm1$. For $\kappa\to0$, these become $\ve \approx (\mu + \mu')\kappa^{-1} + \O(1)$, and since the physically relevant solutions are regular in the decoupled limit, we choose $\mu = -\mu' = \mp 1$. Thus, the interface spectrum is given by 
\begin{equation}
    \ve = \pm \left[ \sqrt{(\eta_y \pm \sinh\lambda)^2 + (1+\eta_z)^2} -\cosh\lambda \right]. 
\end{equation} 
We can now use this expression to compute nodal separation for a fixed energy. 

For homochiral connectivity, the interface spectrum becomes 
\begin{equation*}
    \ve_\text{hom}(k_y) = 
    \pm \left[\sqrt{(\cos k_y - \cos b_0 \pm \sinh\lambda)^2 + \cos^2 b_z} - \cosh\lambda \right]. 
\end{equation*}
For $b_z = \pi/2$, this further simplifies to 
\begin{align}
    \ve_\text{hom}(k_y)  
    = \pm \left[ \sinh\lambda \pm (\cos k_y - \cos b_0)- \cosh\lambda \right]
    = \cos k_y - \cos b_0 \mp \kappa, 
\end{align}
since $\cosh\lambda \pm \sinh\lambda = \e^{\pm\lambda} = \kappa^{\pm 1}$. For $\ve = 0$, we thus get $k_{y,\pm} = \cos^{-1}(\cos b_0 \pm \kappa)$, so that the minimum separation is given by 
\begin{equation}
    \Delta(\kappa) = \cos^{-1}(\cos b_0 - \kappa) - \cos^{-1}(\cos b_0 + \kappa).
\end{equation}

For heterochiral connectivity, we get 
\begin{equation*}
    \ve_\text{het}(k_z) = 
    \pm \left[ \sqrt{(\sin k_z \pm \sinh\lambda)^2 + (1+\cos b_z - \cos k_z)^2} -\cosh\lambda \right].
\end{equation*}
For $b_z = \pi/2$, this becomes  
\begin{align}
    \ve_\text{het}(k_z) = 
    \pm \left[ \sqrt{\cosh^2\lambda - 2 \cosh\lambda \cos(k_z \pm \varphi) + 1} - \cosh\lambda \right], 
\end{align}
where $\varphi = \tan^{-1}(\sinh\lambda)$. For $\ve = 0$, we get $\cos(k_z \pm \varphi) = \frac12 \cos\phi$. Since $\varphi\to\pi/2$ as $\kappa\to0$, the two solutions closest to $k_z = 0$ are given by $k_{z,\pm} = \pm \left[ \cos^{-1}\left( \frac12 \cos \varphi \right) - \phi \right]$, so that the minimum separation is given by 
\begin{equation}
    \Delta(\kappa) 
    = 2 \left[ \cos^{-1}\left( \frac12 \cos \varphi \right) - \varphi \right]
    = 2 \left[ \cos^{-1}\left( \frac{\kappa}{1+\kappa^2} \right) - \tan^{-1} \left( \frac{1-\kappa^2}{2\kappa} \right) \right].
\end{equation}
\end{widetext}

\bibliography{library}	

\begin{thebibliography}{43}%
\makeatletter
\providecommand \@ifxundefined [1]{%
 \@ifx{#1\undefined}
}%
\providecommand \@ifnum [1]{%
 \ifnum #1\expandafter \@firstoftwo
 \else \expandafter \@secondoftwo
 \fi
}%
\providecommand \@ifx [1]{%
 \ifx #1\expandafter \@firstoftwo
 \else \expandafter \@secondoftwo
 \fi
}%
\providecommand \natexlab [1]{#1}%
\providecommand \enquote  [1]{``#1''}%
\providecommand \bibnamefont  [1]{#1}%
\providecommand \bibfnamefont [1]{#1}%
\providecommand \citenamefont [1]{#1}%
\providecommand \href@noop [0]{\@secondoftwo}%
\providecommand \href [0]{\begingroup \@sanitize@url \@href}%
\providecommand \@href[1]{\@@startlink{#1}\@@href}%
\providecommand \@@href[1]{\endgroup#1\@@endlink}%
\providecommand \@sanitize@url [0]{\catcode `\\12\catcode `\$12\catcode
  `\&12\catcode `\#12\catcode `\^12\catcode `\_12\catcode `\%12\relax}%
\providecommand \@@startlink[1]{}%
\providecommand \@@endlink[0]{}%
\providecommand \url  [0]{\begingroup\@sanitize@url \@url }%
\providecommand \@url [1]{\endgroup\@href {#1}{\urlprefix }}%
\providecommand \urlprefix  [0]{URL }%
\providecommand \Eprint [0]{\href }%
\providecommand \doibase [0]{https://doi.org/}%
\providecommand \selectlanguage [0]{\@gobble}%
\providecommand \bibinfo  [0]{\@secondoftwo}%
\providecommand \bibfield  [0]{\@secondoftwo}%
\providecommand \translation [1]{[#1]}%
\providecommand \BibitemOpen [0]{}%
\providecommand \bibitemStop [0]{}%
\providecommand \bibitemNoStop [0]{.\EOS\space}%
\providecommand \EOS [0]{\spacefactor3000\relax}%
\providecommand \BibitemShut  [1]{\csname bibitem#1\endcsname}%
\let\auto@bib@innerbib\@empty
\bibitem [{\citenamefont {Wan}\ \emph {et~al.}(2011)\citenamefont {Wan},
  \citenamefont {Turner}, \citenamefont {Vishwanath},\ and\ \citenamefont
  {Savrasov}}]{Wan2011}%
  \BibitemOpen
  \bibfield  {author} {\bibinfo {author} {\bibfnamefont {X.}~\bibnamefont
  {Wan}}, \bibinfo {author} {\bibfnamefont {A.~M.}\ \bibnamefont {Turner}},
  \bibinfo {author} {\bibfnamefont {A.}~\bibnamefont {Vishwanath}},\ and\
  \bibinfo {author} {\bibfnamefont {S.~Y.}\ \bibnamefont {Savrasov}},\
  }\bibfield  {title} {\bibinfo {title} {{Topological semimetal and Fermi-arc
  surface states in the electronic structure of pyrochlore iridates}},\ }\href
  {https://doi.org/10.1103/PhysRevB.83.205101} {\bibfield  {journal} {\bibinfo
  {journal} {Phys. Rev. B}\ }\textbf {\bibinfo {volume} {83}},\ \bibinfo
  {pages} {205101} (\bibinfo {year} {2011})}\BibitemShut {NoStop}%
\bibitem [{\citenamefont {Burkov}\ and\ \citenamefont
  {Balents}(2011)}]{Burkov2011}%
  \BibitemOpen
  \bibfield  {author} {\bibinfo {author} {\bibfnamefont {A.~A.}\ \bibnamefont
  {Burkov}}\ and\ \bibinfo {author} {\bibfnamefont {L.}~\bibnamefont
  {Balents}},\ }\bibfield  {title} {\bibinfo {title} {{Weyl Semimetal in a
  Topological Insulator Multilayer}},\ }\href
  {https://doi.org/https://doi.org/10.1103/PhysRevLett.107.127205} {\bibfield
  {journal} {\bibinfo  {journal} {Phys. Rev. Lett}\ }\textbf {\bibinfo {volume}
  {107}},\ \bibinfo {pages} {127205} (\bibinfo {year} {2011})}\BibitemShut
  {NoStop}%
\bibitem [{\citenamefont {Xu}\ \emph {et~al.}(2011)\citenamefont {Xu},
  \citenamefont {Weng}, \citenamefont {Wang}, \citenamefont {Dai},\ and\
  \citenamefont {Fang}}]{Xu2011}%
  \BibitemOpen
  \bibfield  {author} {\bibinfo {author} {\bibfnamefont {G.}~\bibnamefont
  {Xu}}, \bibinfo {author} {\bibfnamefont {H.}~\bibnamefont {Weng}}, \bibinfo
  {author} {\bibfnamefont {Z.}~\bibnamefont {Wang}}, \bibinfo {author}
  {\bibfnamefont {X.}~\bibnamefont {Dai}},\ and\ \bibinfo {author}
  {\bibfnamefont {Z.}~\bibnamefont {Fang}},\ }\bibfield  {title} {\bibinfo
  {title} {{Chern Semimetal and the Quantized Anomalous Hall Effect in
  ${\mathrm{HgCr}}_{2}{\mathrm{Se}}_{4}$}},\ }\href
  {https://doi.org/10.1103/PhysRevLett.107.186806} {\bibfield  {journal}
  {\bibinfo  {journal} {Phys. Rev. Lett.}\ }\textbf {\bibinfo {volume} {107}},\
  \bibinfo {pages} {186806} (\bibinfo {year} {2011})}\BibitemShut {NoStop}%
\bibitem [{\citenamefont {Xu}\ \emph {et~al.}(2015{\natexlab{a}})\citenamefont
  {Xu}, \citenamefont {Liu}, \citenamefont {Kushwaha}, \citenamefont {Sankar},
  \citenamefont {Krizan}, \citenamefont {Belopolski}, \citenamefont {Neupane},
  \citenamefont {Bian}, \citenamefont {Alidoust}, \citenamefont {Chang},
  \citenamefont {Jeng}, \citenamefont {Huang}, \citenamefont {Tsai},
  \citenamefont {Lin}, \citenamefont {Shibayev}, \citenamefont {Chou},
  \citenamefont {Cava},\ and\ \citenamefont {Hasan}}]{Xu2015}%
  \BibitemOpen
  \bibfield  {author} {\bibinfo {author} {\bibfnamefont {S.-Y.}\ \bibnamefont
  {Xu}}, \bibinfo {author} {\bibfnamefont {C.}~\bibnamefont {Liu}}, \bibinfo
  {author} {\bibfnamefont {S.~K.}\ \bibnamefont {Kushwaha}}, \bibinfo {author}
  {\bibfnamefont {R.}~\bibnamefont {Sankar}}, \bibinfo {author} {\bibfnamefont
  {J.~W.}\ \bibnamefont {Krizan}}, \bibinfo {author} {\bibfnamefont
  {I.}~\bibnamefont {Belopolski}}, \bibinfo {author} {\bibfnamefont
  {M.}~\bibnamefont {Neupane}}, \bibinfo {author} {\bibfnamefont
  {G.}~\bibnamefont {Bian}}, \bibinfo {author} {\bibfnamefont {N.}~\bibnamefont
  {Alidoust}}, \bibinfo {author} {\bibfnamefont {T.-R.}\ \bibnamefont {Chang}},
  \bibinfo {author} {\bibfnamefont {H.-T.}\ \bibnamefont {Jeng}}, \bibinfo
  {author} {\bibfnamefont {C.-Y.}\ \bibnamefont {Huang}}, \bibinfo {author}
  {\bibfnamefont {W.-F.}\ \bibnamefont {Tsai}}, \bibinfo {author}
  {\bibfnamefont {H.}~\bibnamefont {Lin}}, \bibinfo {author} {\bibfnamefont
  {P.~P.}\ \bibnamefont {Shibayev}}, \bibinfo {author} {\bibfnamefont {F.-C.}\
  \bibnamefont {Chou}}, \bibinfo {author} {\bibfnamefont {R.~J.}\ \bibnamefont
  {Cava}},\ and\ \bibinfo {author} {\bibfnamefont {M.~Z.}\ \bibnamefont
  {Hasan}},\ }\bibfield  {title} {\bibinfo {title} {{Observation of Fermi arc
  surface states in a topological metal}},\ }\href {https://doi.org/DOI:
  10.1126/science.1256742} {\bibfield  {journal} {\bibinfo  {journal}
  {Science}\ }\textbf {\bibinfo {volume} {347}},\ \bibinfo {pages} {294}
  (\bibinfo {year} {2015}{\natexlab{a}})}\BibitemShut {NoStop}%
\bibitem [{\citenamefont {Xu}\ \emph {et~al.}(2015{\natexlab{b}})\citenamefont
  {Xu}, \citenamefont {Alidoust}, \citenamefont {Belopolski}, \citenamefont
  {Yuan}, \citenamefont {Bian}, \citenamefont {Chang}, \citenamefont {Zheng},
  \citenamefont {Strocov}, \citenamefont {Sanchez}, \citenamefont {Chang},
  \citenamefont {Zhang}, \citenamefont {Mou}, \citenamefont {Wu}, \citenamefont
  {Huang}, \citenamefont {Lee}, \citenamefont {Huang}, \citenamefont {Wang},
  \citenamefont {Bansil}, \citenamefont {Jeng}, \citenamefont {Neupert},
  \citenamefont {Kaminski}, \citenamefont {Lin}, \citenamefont {Jia},\ and\
  \citenamefont {{Zahid Hasan}}}]{Xu2015b}%
  \BibitemOpen
  \bibfield  {author} {\bibinfo {author} {\bibfnamefont {S.-Y.}\ \bibnamefont
  {Xu}}, \bibinfo {author} {\bibfnamefont {N.}~\bibnamefont {Alidoust}},
  \bibinfo {author} {\bibfnamefont {I.}~\bibnamefont {Belopolski}}, \bibinfo
  {author} {\bibfnamefont {Z.}~\bibnamefont {Yuan}}, \bibinfo {author}
  {\bibfnamefont {G.}~\bibnamefont {Bian}}, \bibinfo {author} {\bibfnamefont
  {T.-R.}\ \bibnamefont {Chang}}, \bibinfo {author} {\bibfnamefont
  {H.}~\bibnamefont {Zheng}}, \bibinfo {author} {\bibfnamefont {V.~N.}\
  \bibnamefont {Strocov}}, \bibinfo {author} {\bibfnamefont {D.~S.}\
  \bibnamefont {Sanchez}}, \bibinfo {author} {\bibfnamefont {G.}~\bibnamefont
  {Chang}}, \bibinfo {author} {\bibfnamefont {C.}~\bibnamefont {Zhang}},
  \bibinfo {author} {\bibfnamefont {D.}~\bibnamefont {Mou}}, \bibinfo {author}
  {\bibfnamefont {Y.}~\bibnamefont {Wu}}, \bibinfo {author} {\bibfnamefont
  {L.}~\bibnamefont {Huang}}, \bibinfo {author} {\bibfnamefont {C.-C.}\
  \bibnamefont {Lee}}, \bibinfo {author} {\bibfnamefont {S.-M.}\ \bibnamefont
  {Huang}}, \bibinfo {author} {\bibfnamefont {B.}~\bibnamefont {Wang}},
  \bibinfo {author} {\bibfnamefont {A.}~\bibnamefont {Bansil}}, \bibinfo
  {author} {\bibfnamefont {H.-T.}\ \bibnamefont {Jeng}}, \bibinfo {author}
  {\bibfnamefont {T.}~\bibnamefont {Neupert}}, \bibinfo {author} {\bibfnamefont
  {A.}~\bibnamefont {Kaminski}}, \bibinfo {author} {\bibfnamefont
  {H.}~\bibnamefont {Lin}}, \bibinfo {author} {\bibfnamefont {S.}~\bibnamefont
  {Jia}},\ and\ \bibinfo {author} {\bibfnamefont {M.}~\bibnamefont {{Zahid
  Hasan}}},\ }\bibfield  {title} {\bibinfo {title} {{Discovery of a Weyl
  fermion state with Fermi arcs in niobium arsenide}},\ }\href
  {https://doi.org/10.1038/nphys3437} {\bibfield  {journal} {\bibinfo
  {journal} {Nat. Phys.}\ }\textbf {\bibinfo {volume} {11}},\ \bibinfo {pages}
  {748} (\bibinfo {year} {2015}{\natexlab{b}})}\BibitemShut {NoStop}%
\bibitem [{\citenamefont {Lv}\ \emph {et~al.}(2015)\citenamefont {Lv},
  \citenamefont {Weng}, \citenamefont {Fu}, \citenamefont {Wang}, \citenamefont
  {Miao}, \citenamefont {Ma}, \citenamefont {Richard}, \citenamefont {Huang},
  \citenamefont {Zhao}, \citenamefont {Chen}, \citenamefont {Fang},
  \citenamefont {Dai}, \citenamefont {Qian},\ and\ \citenamefont
  {Ding}}]{Lv2015}%
  \BibitemOpen
  \bibfield  {author} {\bibinfo {author} {\bibfnamefont {B.~Q.}\ \bibnamefont
  {Lv}}, \bibinfo {author} {\bibfnamefont {H.~M.}\ \bibnamefont {Weng}},
  \bibinfo {author} {\bibfnamefont {B.~B.}\ \bibnamefont {Fu}}, \bibinfo
  {author} {\bibfnamefont {X.~P.}\ \bibnamefont {Wang}}, \bibinfo {author}
  {\bibfnamefont {H.}~\bibnamefont {Miao}}, \bibinfo {author} {\bibfnamefont
  {J.}~\bibnamefont {Ma}}, \bibinfo {author} {\bibfnamefont {P.}~\bibnamefont
  {Richard}}, \bibinfo {author} {\bibfnamefont {X.~C.}\ \bibnamefont {Huang}},
  \bibinfo {author} {\bibfnamefont {L.~X.}\ \bibnamefont {Zhao}}, \bibinfo
  {author} {\bibfnamefont {G.~F.}\ \bibnamefont {Chen}}, \bibinfo {author}
  {\bibfnamefont {Z.}~\bibnamefont {Fang}}, \bibinfo {author} {\bibfnamefont
  {X.}~\bibnamefont {Dai}}, \bibinfo {author} {\bibfnamefont {T.}~\bibnamefont
  {Qian}},\ and\ \bibinfo {author} {\bibfnamefont {H.}~\bibnamefont {Ding}},\
  }\bibfield  {title} {\bibinfo {title} {{Experimental discovery of weyl
  semimetal TaAs}},\ }\href {https://doi.org/10.1103/PhysRevX.5.031013}
  {\bibfield  {journal} {\bibinfo  {journal} {Phys. Rev. X}\ }\textbf {\bibinfo
  {volume} {5}},\ \bibinfo {pages} {031013} (\bibinfo {year}
  {2015})}\BibitemShut {NoStop}%
\bibitem [{\citenamefont {Armitage}\ \emph {et~al.}(2018)\citenamefont
  {Armitage}, \citenamefont {Mele},\ and\ \citenamefont
  {Vishwanath}}]{Armitage2017}%
  \BibitemOpen
  \bibfield  {author} {\bibinfo {author} {\bibfnamefont {N.~P.}\ \bibnamefont
  {Armitage}}, \bibinfo {author} {\bibfnamefont {E.~J.}\ \bibnamefont {Mele}},\
  and\ \bibinfo {author} {\bibfnamefont {A.}~\bibnamefont {Vishwanath}},\
  }\bibfield  {title} {\bibinfo {title} {{Weyl and Dirac Semimetals in Three
  Dimensional Solids}},\ }\href {https://doi.org/10.1103/RevModPhys.90.015001}
  {\bibfield  {journal} {\bibinfo  {journal} {Rev. Mod. Phys.}\ }\textbf
  {\bibinfo {volume} {90}},\ \bibinfo {pages} {015001} (\bibinfo {year}
  {2018})}\BibitemShut {NoStop}%
\bibitem [{\citenamefont {Yan}\ and\ \citenamefont {Felser}(2017)}]{Yan2017}%
  \BibitemOpen
  \bibfield  {author} {\bibinfo {author} {\bibfnamefont {B.}~\bibnamefont
  {Yan}}\ and\ \bibinfo {author} {\bibfnamefont {C.}~\bibnamefont {Felser}},\
  }\bibfield  {title} {\bibinfo {title} {{Topological Materials: Weyl
  Semimetals}},\ }\href
  {https://doi.org/10.1146/annurev-conmatphys-031016-025458} {\bibfield
  {journal} {\bibinfo  {journal} {Annu. Rev. Condens. Matter Phys.}\ }\textbf
  {\bibinfo {volume} {8}},\ \bibinfo {pages} {337} (\bibinfo {year}
  {2017})}\BibitemShut {NoStop}%
\bibitem [{\citenamefont {Burkov}(2018)}]{Burkov2017}%
  \BibitemOpen
  \bibfield  {author} {\bibinfo {author} {\bibfnamefont {A.~A.}\ \bibnamefont
  {Burkov}},\ }\bibfield  {title} {\bibinfo {title} {{Weyl Metals}},\ }\href
  {https://doi.org/10.1146/annurev-conmatphys-033117-054129} {\bibfield
  {journal} {\bibinfo  {journal} {Annu. Rev. Condens. Matter Phys.}\ }\textbf
  {\bibinfo {volume} {9}},\ \bibinfo {pages} {359} (\bibinfo {year}
  {2018})}\BibitemShut {NoStop}%
\bibitem [{\citenamefont {Hasan}\ \emph {et~al.}(2021)\citenamefont {Hasan},
  \citenamefont {Chang}, \citenamefont {Belopolski}, \citenamefont {Bian},
  \citenamefont {Xu},\ and\ \citenamefont {Yin}}]{Hasan2021}%
  \BibitemOpen
  \bibfield  {author} {\bibinfo {author} {\bibfnamefont {M.~Z.}\ \bibnamefont
  {Hasan}}, \bibinfo {author} {\bibfnamefont {G.}~\bibnamefont {Chang}},
  \bibinfo {author} {\bibfnamefont {I.}~\bibnamefont {Belopolski}}, \bibinfo
  {author} {\bibfnamefont {G.}~\bibnamefont {Bian}}, \bibinfo {author}
  {\bibfnamefont {S.~Y.}\ \bibnamefont {Xu}},\ and\ \bibinfo {author}
  {\bibfnamefont {J.~X.}\ \bibnamefont {Yin}},\ }\bibfield  {title} {\bibinfo
  {title} {{Weyl, Dirac and high-fold chiral fermions in topological quantum
  matter}},\ }\href {https://doi.org/10.1038/s41578-021-00301-3} {\bibfield
  {journal} {\bibinfo  {journal} {Nat. Rev. Mater.}\ }\textbf {\bibinfo
  {volume} {6}},\ \bibinfo {pages} {784} (\bibinfo {year} {2021})}\BibitemShut
  {NoStop}%
\bibitem [{\citenamefont {Bernevig}\ \emph {et~al.}(2022)\citenamefont
  {Bernevig}, \citenamefont {Felser},\ and\ \citenamefont
  {Beidenkopf}}]{Bernevig2022a}%
  \BibitemOpen
  \bibfield  {author} {\bibinfo {author} {\bibfnamefont {B.~A.}\ \bibnamefont
  {Bernevig}}, \bibinfo {author} {\bibfnamefont {C.}~\bibnamefont {Felser}},\
  and\ \bibinfo {author} {\bibfnamefont {H.}~\bibnamefont {Beidenkopf}},\
  }\bibfield  {title} {\bibinfo {title} {{Progress and prospects in magnetic
  topological materials}},\ }\href {https://doi.org/10.1038/s41586-021-04105-x}
  {\bibfield  {journal} {\bibinfo  {journal} {Nature}\ }\textbf {\bibinfo
  {volume} {603}},\ \bibinfo {pages} {41} (\bibinfo {year} {2022})}\BibitemShut
  {NoStop}%
\bibitem [{\citenamefont {Adler}(1969)}]{Adler1969}%
  \BibitemOpen
  \bibfield  {author} {\bibinfo {author} {\bibfnamefont {S.~L.}\ \bibnamefont
  {Adler}},\ }\bibfield  {title} {\bibinfo {title} {{Axial-Vector Vertex in
  Spinor Electrodynamics}},\ }\href@noop {} {\bibfield  {journal} {\bibinfo
  {journal} {Phys. Rev.}\ }\textbf {\bibinfo {volume} {177(5)}},\ \bibinfo
  {pages} {2426} (\bibinfo {year} {1969})}\BibitemShut {NoStop}%
\bibitem [{\citenamefont {Bell}\ and\ \citenamefont {Jackiw}(1969)}]{Bell1969}%
  \BibitemOpen
  \bibfield  {author} {\bibinfo {author} {\bibfnamefont {J.~S.}\ \bibnamefont
  {Bell}}\ and\ \bibinfo {author} {\bibfnamefont {R.~W.}\ \bibnamefont
  {Jackiw}},\ }\bibfield  {title} {\bibinfo {title} {{E {$\pi$}}},\ }\href@noop
  {} {\bibfield  {journal} {\bibinfo  {journal} {Nuovo Cim.}\ }\textbf
  {\bibinfo {volume} {60}},\ \bibinfo {pages} {47} (\bibinfo {year}
  {1969})}\BibitemShut {NoStop}%
\bibitem [{\citenamefont {Nielsen}\ and\ \citenamefont
  {Ninomiya}(1983)}]{Nielsen1983}%
  \BibitemOpen
  \bibfield  {author} {\bibinfo {author} {\bibfnamefont {H.~B.}\ \bibnamefont
  {Nielsen}}\ and\ \bibinfo {author} {\bibfnamefont {M.}~\bibnamefont
  {Ninomiya}},\ }\bibfield  {title} {\bibinfo {title} {{The Adler-Bell-Jackiw
  anomaly and Weyl fermions in a crystal}},\ }\href@noop {} {\bibfield
  {journal} {\bibinfo  {journal} {Phys. Lett. B}\ }\textbf {\bibinfo {volume}
  {130(6)}},\ \bibinfo {pages} {389} (\bibinfo {year} {1983})}\BibitemShut
  {NoStop}%
\bibitem [{\citenamefont {Dwivedi}(2018)}]{Dwivedi2018}%
  \BibitemOpen
  \bibfield  {author} {\bibinfo {author} {\bibfnamefont {V.}~\bibnamefont
  {Dwivedi}},\ }\bibfield  {title} {\bibinfo {title} {{Fermi arc reconstruction
  at junctions between Weyl semimetals}},\ }\href
  {https://doi.org/10.1103/PhysRevB.97.064201} {\bibfield  {journal} {\bibinfo
  {journal} {Phys. Rev. B}\ }\textbf {\bibinfo {volume} {97}},\ \bibinfo
  {pages} {064201} (\bibinfo {year} {2018})}\BibitemShut {NoStop}%
\bibitem [{\citenamefont {Abdulla}\ \emph {et~al.}(2021)\citenamefont
  {Abdulla}, \citenamefont {Rao},\ and\ \citenamefont {Murthy}}]{Abdulla2021}%
  \BibitemOpen
  \bibfield  {author} {\bibinfo {author} {\bibfnamefont {F.}~\bibnamefont
  {Abdulla}}, \bibinfo {author} {\bibfnamefont {S.}~\bibnamefont {Rao}},\ and\
  \bibinfo {author} {\bibfnamefont {G.}~\bibnamefont {Murthy}},\ }\bibfield
  {title} {\bibinfo {title} {{Fermi arc reconstruction at the interface of
  twisted Weyl semimetals}},\ }\href
  {https://doi.org/10.1103/PhysRevB.103.235308} {\bibfield  {journal} {\bibinfo
   {journal} {Phys. Rev. B}\ }\textbf {\bibinfo {volume} {103}},\ \bibinfo
  {pages} {235308} (\bibinfo {year} {2021})}\BibitemShut {NoStop}%
\bibitem [{\citenamefont {Mathur}\ \emph {et~al.}(2023)\citenamefont {Mathur},
  \citenamefont {Yuan}, \citenamefont {Cheng}, \citenamefont {Kaushik},
  \citenamefont {Robredo},\ and\ \citenamefont {Maia}}]{Mathur2022}%
  \BibitemOpen
  \bibfield  {author} {\bibinfo {author} {\bibfnamefont {N.}~\bibnamefont
  {Mathur}}, \bibinfo {author} {\bibfnamefont {F.}~\bibnamefont {Yuan}},
  \bibinfo {author} {\bibfnamefont {G.}~\bibnamefont {Cheng}}, \bibinfo
  {author} {\bibfnamefont {S.}~\bibnamefont {Kaushik}}, \bibinfo {author}
  {\bibfnamefont {I.}~\bibnamefont {Robredo}},\ and\ \bibinfo {author}
  {\bibfnamefont {G.}~\bibnamefont {Maia}},\ }\bibfield  {title} {\bibinfo
  {title} {{Atomically Sharp Internal Interface in a Chiral Weyl Semimetal
  Nanowire}},\ }\href {https://doi.org/10.1021/acs.nanolett.2c05100} {\bibfield
   {journal} {\bibinfo  {journal} {Nano Lett.}\ }\textbf {\bibinfo {volume}
  {23}},\ \bibinfo {pages} {2695} (\bibinfo {year} {2023})}\BibitemShut
  {NoStop}%
\bibitem [{\citenamefont {Kaushik}\ \emph {et~al.}()\citenamefont {Kaushik},
  \citenamefont {Robredo}, \citenamefont {Mathur}, \citenamefont {Schoop},
  \citenamefont {Jin}, \citenamefont {Vergniory},\ and\ \citenamefont
  {Cano}}]{Kaushik2022}%
  \BibitemOpen
  \bibfield  {author} {\bibinfo {author} {\bibfnamefont {S.}~\bibnamefont
  {Kaushik}}, \bibinfo {author} {\bibfnamefont {I.}~\bibnamefont {Robredo}},
  \bibinfo {author} {\bibfnamefont {N.}~\bibnamefont {Mathur}}, \bibinfo
  {author} {\bibfnamefont {L.~M.}\ \bibnamefont {Schoop}}, \bibinfo {author}
  {\bibfnamefont {S.}~\bibnamefont {Jin}}, \bibinfo {author} {\bibfnamefont
  {M.~G.}\ \bibnamefont {Vergniory}},\ and\ \bibinfo {author} {\bibfnamefont
  {J.}~\bibnamefont {Cano}},\ }\bibfield  {title} {\bibinfo {title} {{Transport
  signatures of Fermi arcs at twin boundaries in Weyl materials}},\ }\href
  {http://arxiv.org/abs/2207.14109} {\bibfield  {journal} {\bibinfo  {journal}
  {arXiv}\ }}\Eprint {https://arxiv.org/abs/2207.14109} {arXiv:2207.14109}
  \BibitemShut {NoStop}%
\bibitem [{\citenamefont {Murthy}\ \emph {et~al.}(2020)\citenamefont {Murthy},
  \citenamefont {Fertig},\ and\ \citenamefont {Shimshoni}}]{Murthy2020}%
  \BibitemOpen
  \bibfield  {author} {\bibinfo {author} {\bibfnamefont {G.}~\bibnamefont
  {Murthy}}, \bibinfo {author} {\bibfnamefont {H.~A.}\ \bibnamefont {Fertig}},\
  and\ \bibinfo {author} {\bibfnamefont {E.}~\bibnamefont {Shimshoni}},\
  }\bibfield  {title} {\bibinfo {title} {{Surface states and arcless angles in
  twisted Weyl semimetals}},\ }\href
  {https://doi.org/10.1103/PhysRevResearch.2.013367} {\bibfield  {journal}
  {\bibinfo  {journal} {Phys. Rev. Res.}\ }\textbf {\bibinfo {volume} {2}},\
  \bibinfo {pages} {13367} (\bibinfo {year} {2020})}\BibitemShut {NoStop}%
\bibitem [{\citenamefont {Kundu}\ \emph {et~al.}(2023)\citenamefont {Kundu},
  \citenamefont {Fertig},\ and\ \citenamefont {Kundu}}]{Kundu2023}%
  \BibitemOpen
  \bibfield  {author} {\bibinfo {author} {\bibfnamefont {R.}~\bibnamefont
  {Kundu}}, \bibinfo {author} {\bibfnamefont {H.~A.}\ \bibnamefont {Fertig}},\
  and\ \bibinfo {author} {\bibfnamefont {A.}~\bibnamefont {Kundu}},\ }\bibfield
   {title} {\bibinfo {title} {{Broken symmetry and competing orders in Weyl
  semimetal interfaces}},\ }\href
  {https://doi.org/10.1103/PhysRevB.107.L041402} {\bibfield  {journal}
  {\bibinfo  {journal} {Phys. Rev. B}\ }\textbf {\bibinfo {volume} {107}},\
  \bibinfo {pages} {L041402} (\bibinfo {year} {2023})}\BibitemShut {NoStop}%
\bibitem [{\citenamefont {Chaou}\ \emph {et~al.}(2023)\citenamefont {Chaou},
  \citenamefont {Dwivedi},\ and\ \citenamefont {Breitkreiz}}]{Chaou2023}%
  \BibitemOpen
  \bibfield  {author} {\bibinfo {author} {\bibfnamefont {A.~Y.}\ \bibnamefont
  {Chaou}}, \bibinfo {author} {\bibfnamefont {V.}~\bibnamefont {Dwivedi}},\
  and\ \bibinfo {author} {\bibfnamefont {M.}~\bibnamefont {Breitkreiz}},\
  }\bibfield  {title} {\bibinfo {title} {{Magnetic Breakdown and Chiral
  Magnetic Effect at Weyl-Semimetal Tunnel Junctions}},\ }\href
  {https://doi.org/10.1103/PhysRevB.107.L241109} {\bibfield  {journal}
  {\bibinfo  {journal} {Phys. Rev. B}\ }\textbf {\bibinfo {volume} {107}},\
  \bibinfo {pages} {L241109} (\bibinfo {year} {2023})}\BibitemShut {NoStop}%
\bibitem [{\citenamefont {Parameswaran}\ \emph {et~al.}(2014)\citenamefont
  {Parameswaran}, \citenamefont {Grover}, \citenamefont {Abanin}, \citenamefont
  {Pesin},\ and\ \citenamefont {Vishwanath}}]{Parameswaran2014}%
  \BibitemOpen
  \bibfield  {author} {\bibinfo {author} {\bibfnamefont {S.~A.}\ \bibnamefont
  {Parameswaran}}, \bibinfo {author} {\bibfnamefont {T.}~\bibnamefont
  {Grover}}, \bibinfo {author} {\bibfnamefont {D.~A.}\ \bibnamefont {Abanin}},
  \bibinfo {author} {\bibfnamefont {D.~A.}\ \bibnamefont {Pesin}},\ and\
  \bibinfo {author} {\bibfnamefont {A.}~\bibnamefont {Vishwanath}},\ }\bibfield
   {title} {\bibinfo {title} {{Probing the Chiral Anomaly with Nonlocal
  Transport in Three-Dimensional Topological Semimetals}},\ }\href
  {https://doi.org/10.1103/PhysRevX.4.031035} {\bibfield  {journal} {\bibinfo
  {journal} {Phys. Rev. X}\ }\textbf {\bibinfo {volume} {4}},\ \bibinfo {pages}
  {031035} (\bibinfo {year} {2014})}\BibitemShut {NoStop}%
\bibitem [{\citenamefont {Baum}\ \emph {et~al.}(2015)\citenamefont {Baum},
  \citenamefont {Berg}, \citenamefont {Parameswaran},\ and\ \citenamefont
  {Stern}}]{Baum2015}%
  \BibitemOpen
  \bibfield  {author} {\bibinfo {author} {\bibfnamefont {Y.}~\bibnamefont
  {Baum}}, \bibinfo {author} {\bibfnamefont {E.}~\bibnamefont {Berg}}, \bibinfo
  {author} {\bibfnamefont {S.~A.}\ \bibnamefont {Parameswaran}},\ and\ \bibinfo
  {author} {\bibfnamefont {A.}~\bibnamefont {Stern}},\ }\bibfield  {title}
  {\bibinfo {title} {{Current at a Distance and Resonant Transparency in Weyl
  Semimetals}},\ }\href {https://doi.org/10.1103/PhysRevX.5.041046} {\bibfield
  {journal} {\bibinfo  {journal} {Phys. Rev. X}\ }\textbf {\bibinfo {volume}
  {5}},\ \bibinfo {pages} {041046} (\bibinfo {year} {2015})}\BibitemShut
  {NoStop}%
\bibitem [{\citenamefont {Shoenberg}(1984)}]{Shoenberg1984}%
  \BibitemOpen
  \bibfield  {author} {\bibinfo {author} {\bibfnamefont {D.}~\bibnamefont
  {Shoenberg}},\ }\href {https://doi.org/10.1017/CBO9780511897870} {\emph
  {\bibinfo {title} {{Magnetic Oscillations in Metals}}}},\ Cambridge
  Monographs on Physics\ (\bibinfo  {publisher} {Cambridge University Press},\
  \bibinfo {year} {1984})\BibitemShut {NoStop}%
\bibitem [{\citenamefont {Potter}\ \emph {et~al.}(2014)\citenamefont {Potter},
  \citenamefont {Kimchi},\ and\ \citenamefont {Vishwanath}}]{Potter2014}%
  \BibitemOpen
  \bibfield  {author} {\bibinfo {author} {\bibfnamefont {A.}~\bibnamefont
  {Potter}}, \bibinfo {author} {\bibfnamefont {I.}~\bibnamefont {Kimchi}},\
  and\ \bibinfo {author} {\bibfnamefont {A.}~\bibnamefont {Vishwanath}},\
  }\bibfield  {title} {\bibinfo {title} {{Quantum Oscillations from Surface
  Fermi-Arcs in Weyl and Dirac Semi-Metals}},\ }\href
  {https://doi.org/10.1038/ncomms6161} {\bibfield  {journal} {\bibinfo
  {journal} {Nat. Commun.}\ }\textbf {\bibinfo {volume} {5}},\ \bibinfo {pages}
  {5161} (\bibinfo {year} {2014})}\BibitemShut {NoStop}%
\bibitem [{\citenamefont {Moll}\ \emph {et~al.}(2016)\citenamefont {Moll},
  \citenamefont {Nair}, \citenamefont {Helm}, \citenamefont {Potter},
  \citenamefont {Kimchi}, \citenamefont {Vishwanath},\ and\ \citenamefont
  {Analytis}}]{Moll2016}%
  \BibitemOpen
  \bibfield  {author} {\bibinfo {author} {\bibfnamefont {P.~J.~W.}\
  \bibnamefont {Moll}}, \bibinfo {author} {\bibfnamefont {N.~L.}\ \bibnamefont
  {Nair}}, \bibinfo {author} {\bibfnamefont {T.}~\bibnamefont {Helm}}, \bibinfo
  {author} {\bibfnamefont {A.~C.}\ \bibnamefont {Potter}}, \bibinfo {author}
  {\bibfnamefont {I.}~\bibnamefont {Kimchi}}, \bibinfo {author} {\bibfnamefont
  {A.}~\bibnamefont {Vishwanath}},\ and\ \bibinfo {author} {\bibfnamefont
  {J.~G.}\ \bibnamefont {Analytis}},\ }\bibfield  {title} {\bibinfo {title}
  {{Transport evidence for Fermi-arc-mediated chirality transfer in the Dirac
  semimetal Cd 3 As 2}},\ }\href {https://doi.org/10.1038/nature18276}
  {\bibfield  {journal} {\bibinfo  {journal} {Nat. Lett.}\ }\textbf {\bibinfo
  {volume} {535}},\ \bibinfo {pages} {266} (\bibinfo {year}
  {2016})}\BibitemShut {NoStop}%
\bibitem [{\citenamefont {Galletti}\ \emph {et~al.}(2019)\citenamefont
  {Galletti}, \citenamefont {Schumann}, \citenamefont {Kealhofer},
  \citenamefont {Goyal},\ and\ \citenamefont {Stemmer}}]{Galletti2019}%
  \BibitemOpen
  \bibfield  {author} {\bibinfo {author} {\bibfnamefont {L.}~\bibnamefont
  {Galletti}}, \bibinfo {author} {\bibfnamefont {T.}~\bibnamefont {Schumann}},
  \bibinfo {author} {\bibfnamefont {D.~A.}\ \bibnamefont {Kealhofer}}, \bibinfo
  {author} {\bibfnamefont {M.}~\bibnamefont {Goyal}},\ and\ \bibinfo {author}
  {\bibfnamefont {S.}~\bibnamefont {Stemmer}},\ }\bibfield  {title} {\bibinfo
  {title} {{Absence of signatures of Weyl orbits in the thickness dependence of
  quantum transport in cadmium arsenide}},\ }\href
  {https://doi.org/10.1103/PhysRevB.99.201401} {\bibfield  {journal} {\bibinfo
  {journal} {Phys. Rev. B}\ }\textbf {\bibinfo {volume} {99}},\ \bibinfo
  {pages} {201401} (\bibinfo {year} {2019})}\BibitemShut {NoStop}%
\bibitem [{\citenamefont {Kar}\ \emph {et~al.}(2023)\citenamefont {Kar},
  \citenamefont {Singh}, \citenamefont {Hsu}, \citenamefont {Lin},
  \citenamefont {Das}, \citenamefont {Cheng}, \citenamefont {Berben},
  \citenamefont {Yang}, \citenamefont {Lin}, \citenamefont {Hsu}, \citenamefont
  {Wiedmann}, \citenamefont {Lee},\ and\ \citenamefont {Lee}}]{Kar2023}%
  \BibitemOpen
  \bibfield  {author} {\bibinfo {author} {\bibfnamefont {U.}~\bibnamefont
  {Kar}}, \bibinfo {author} {\bibfnamefont {A.~K.}\ \bibnamefont {Singh}},
  \bibinfo {author} {\bibfnamefont {Y.-T.}\ \bibnamefont {Hsu}}, \bibinfo
  {author} {\bibfnamefont {C.-Y.}\ \bibnamefont {Lin}}, \bibinfo {author}
  {\bibfnamefont {B.}~\bibnamefont {Das}}, \bibinfo {author} {\bibfnamefont
  {C.-T.}\ \bibnamefont {Cheng}}, \bibinfo {author} {\bibfnamefont
  {M.}~\bibnamefont {Berben}}, \bibinfo {author} {\bibfnamefont
  {S.}~\bibnamefont {Yang}}, \bibinfo {author} {\bibfnamefont {C.-Y.}\
  \bibnamefont {Lin}}, \bibinfo {author} {\bibfnamefont {C.-H.}\ \bibnamefont
  {Hsu}}, \bibinfo {author} {\bibfnamefont {S.}~\bibnamefont {Wiedmann}},
  \bibinfo {author} {\bibfnamefont {W.-C.}\ \bibnamefont {Lee}},\ and\ \bibinfo
  {author} {\bibfnamefont {W.-L.}\ \bibnamefont {Lee}},\ }\bibfield  {title}
  {\bibinfo {title} {{The thickness dependence of quantum oscillations in
  ferromagnetic Weyl metal SrRuO3}},\ }\href
  {https://doi.org/10.1038/s41535-023-00540-3} {\bibfield  {journal} {\bibinfo
  {journal} {npj Quantum Mater.}\ }\textbf {\bibinfo {volume} {8}},\ \bibinfo
  {pages} {8} (\bibinfo {year} {2023})}\BibitemShut {NoStop}%
\bibitem [{\citenamefont {Xu}\ \emph {et~al.}(2015{\natexlab{c}})\citenamefont
  {Xu}, \citenamefont {Alidoust}, \citenamefont {Belopolski}, \citenamefont
  {Yuan}, \citenamefont {Bian}, \citenamefont {Chang}, \citenamefont {Zheng},
  \citenamefont {Strocov}, \citenamefont {Sanchez}, \citenamefont {Chang},
  \citenamefont {Zhang}, \citenamefont {Mou}, \citenamefont {Wu}, \citenamefont
  {Huang}, \citenamefont {Lee}, \citenamefont {Huang}, \citenamefont {Wang},
  \citenamefont {Bansil}, \citenamefont {Jeng}, \citenamefont {Neupert},
  \citenamefont {Kaminski}, \citenamefont {Lin}, \citenamefont {Jia},\ and\
  \citenamefont {{Zahid Hasan}}}]{Xu2015a}%
  \BibitemOpen
  \bibfield  {author} {\bibinfo {author} {\bibfnamefont {S.-Y.}\ \bibnamefont
  {Xu}}, \bibinfo {author} {\bibfnamefont {N.}~\bibnamefont {Alidoust}},
  \bibinfo {author} {\bibfnamefont {I.}~\bibnamefont {Belopolski}}, \bibinfo
  {author} {\bibfnamefont {Z.}~\bibnamefont {Yuan}}, \bibinfo {author}
  {\bibfnamefont {G.}~\bibnamefont {Bian}}, \bibinfo {author} {\bibfnamefont
  {T.-R.}\ \bibnamefont {Chang}}, \bibinfo {author} {\bibfnamefont
  {H.}~\bibnamefont {Zheng}}, \bibinfo {author} {\bibfnamefont {V.~N.}\
  \bibnamefont {Strocov}}, \bibinfo {author} {\bibfnamefont {D.~S.}\
  \bibnamefont {Sanchez}}, \bibinfo {author} {\bibfnamefont {G.}~\bibnamefont
  {Chang}}, \bibinfo {author} {\bibfnamefont {C.}~\bibnamefont {Zhang}},
  \bibinfo {author} {\bibfnamefont {D.}~\bibnamefont {Mou}}, \bibinfo {author}
  {\bibfnamefont {Y.}~\bibnamefont {Wu}}, \bibinfo {author} {\bibfnamefont
  {L.}~\bibnamefont {Huang}}, \bibinfo {author} {\bibfnamefont {C.-C.}\
  \bibnamefont {Lee}}, \bibinfo {author} {\bibfnamefont {S.-M.}\ \bibnamefont
  {Huang}}, \bibinfo {author} {\bibfnamefont {B.}~\bibnamefont {Wang}},
  \bibinfo {author} {\bibfnamefont {A.}~\bibnamefont {Bansil}}, \bibinfo
  {author} {\bibfnamefont {H.-T.}\ \bibnamefont {Jeng}}, \bibinfo {author}
  {\bibfnamefont {T.}~\bibnamefont {Neupert}}, \bibinfo {author} {\bibfnamefont
  {A.}~\bibnamefont {Kaminski}}, \bibinfo {author} {\bibfnamefont
  {H.}~\bibnamefont {Lin}}, \bibinfo {author} {\bibfnamefont {S.}~\bibnamefont
  {Jia}},\ and\ \bibinfo {author} {\bibfnamefont {M.}~\bibnamefont {{Zahid
  Hasan}}},\ }\bibfield  {title} {\bibinfo {title} {{Discovery of a Weyl
  fermion semimetal and topological Fermi arcs}},\ }\href
  {https://doi.org/10.1038/nphys3437} {\bibfield  {journal} {\bibinfo
  {journal} {Science}\ }\textbf {\bibinfo {volume} {349}},\ \bibinfo {pages}
  {613} (\bibinfo {year} {2015}{\natexlab{c}})}\BibitemShut {NoStop}%
\bibitem [{\citenamefont {Chang}\ \emph {et~al.}(2016)\citenamefont {Chang},
  \citenamefont {Xu}, \citenamefont {Zheng}, \citenamefont {Lee}, \citenamefont
  {Huang}, \citenamefont {Belopolski}, \citenamefont {Sanchez}, \citenamefont
  {Bian}, \citenamefont {Alidoust}, \citenamefont {Chang}, \citenamefont {Hsu},
  \citenamefont {Jeng}, \citenamefont {Bansil}, \citenamefont {Lin},\ and\
  \citenamefont {Hasan}}]{Chang2016}%
  \BibitemOpen
  \bibfield  {author} {\bibinfo {author} {\bibfnamefont {G.}~\bibnamefont
  {Chang}}, \bibinfo {author} {\bibfnamefont {S.-Y.}\ \bibnamefont {Xu}},
  \bibinfo {author} {\bibfnamefont {H.}~\bibnamefont {Zheng}}, \bibinfo
  {author} {\bibfnamefont {C.-C.}\ \bibnamefont {Lee}}, \bibinfo {author}
  {\bibfnamefont {S.-M.}\ \bibnamefont {Huang}}, \bibinfo {author}
  {\bibfnamefont {I.}~\bibnamefont {Belopolski}}, \bibinfo {author}
  {\bibfnamefont {D.~S.}\ \bibnamefont {Sanchez}}, \bibinfo {author}
  {\bibfnamefont {G.}~\bibnamefont {Bian}}, \bibinfo {author} {\bibfnamefont
  {N.}~\bibnamefont {Alidoust}}, \bibinfo {author} {\bibfnamefont {T.-R.}\
  \bibnamefont {Chang}}, \bibinfo {author} {\bibfnamefont {C.-H.}\ \bibnamefont
  {Hsu}}, \bibinfo {author} {\bibfnamefont {H.-T.}\ \bibnamefont {Jeng}},
  \bibinfo {author} {\bibfnamefont {A.}~\bibnamefont {Bansil}}, \bibinfo
  {author} {\bibfnamefont {H.}~\bibnamefont {Lin}},\ and\ \bibinfo {author}
  {\bibfnamefont {M.~Z.}\ \bibnamefont {Hasan}},\ }\bibfield  {title} {\bibinfo
  {title} {{Signatures of Fermi Arcs in the Quasiparticle Interferences of the
  Weyl Semimetals TaAs and NbP}},\ }\href
  {https://doi.org/10.1103/PhysRevLett.116.066601} {\bibfield  {journal}
  {\bibinfo  {journal} {Phys. Rev. Lett.}\ }\textbf {\bibinfo {volume} {116}},\
  \bibinfo {pages} {66601} (\bibinfo {year} {2016})}\BibitemShut {NoStop}%
\bibitem [{\citenamefont {Cohen}\ and\ \citenamefont
  {Falicov}(1961)}]{Cohen1961}%
  \BibitemOpen
  \bibfield  {author} {\bibinfo {author} {\bibfnamefont {M.~H.}\ \bibnamefont
  {Cohen}}\ and\ \bibinfo {author} {\bibfnamefont {L.~M.}\ \bibnamefont
  {Falicov}},\ }\bibfield  {title} {\bibinfo {title} {{Magnetic breakdown in
  crystals}},\ }\href {https://doi.org/10.1103/PhysRevLett.7.231} {\bibfield
  {journal} {\bibinfo  {journal} {Phys. Rev. Lett.}\ }\textbf {\bibinfo
  {volume} {7}},\ \bibinfo {pages} {231} (\bibinfo {year} {1961})}\BibitemShut
  {NoStop}%
\bibitem [{\citenamefont {Blount}(1962)}]{Blount1962}%
  \BibitemOpen
  \bibfield  {author} {\bibinfo {author} {\bibfnamefont {E.~I.}\ \bibnamefont
  {Blount}},\ }\bibfield  {title} {\bibinfo {title} {{Bloch electrons in a
  magnetic field}},\ }\href@noop {} {\bibfield  {journal} {\bibinfo  {journal}
  {Phys. Rev.}\ }\textbf {\bibinfo {volume} {126}},\ \bibinfo {pages} {1636}
  (\bibinfo {year} {1962})}\BibitemShut {NoStop}%
\bibitem [{\citenamefont {Kaganov}\ and\ \citenamefont
  {Slutskin}(1983)}]{Kaganov1983}%
  \BibitemOpen
  \bibfield  {author} {\bibinfo {author} {\bibfnamefont {M.~I.}\ \bibnamefont
  {Kaganov}}\ and\ \bibinfo {author} {\bibfnamefont {A.~A.}\ \bibnamefont
  {Slutskin}},\ }\bibfield  {title} {\bibinfo {title} {{Coherent magnetic
  breakdown}},\ }\href {https://doi.org/10.1016/0370-1573(83)90006-6}
  {\bibfield  {journal} {\bibinfo  {journal} {Phys. Rep.}\ }\textbf {\bibinfo
  {volume} {98}},\ \bibinfo {pages} {189} (\bibinfo {year} {1983})}\BibitemShut
  {NoStop}%
\bibitem [{\citenamefont {Falicov}\ and\ \citenamefont
  {Stachowiak}(1966)}]{Falicov1966}%
  \BibitemOpen
  \bibfield  {author} {\bibinfo {author} {\bibfnamefont {L.~M.}\ \bibnamefont
  {Falicov}}\ and\ \bibinfo {author} {\bibfnamefont {H.}~\bibnamefont
  {Stachowiak}},\ }\bibfield  {title} {\bibinfo {title} {{Theory of the de
  Haas-van Alphen effect in a system of coupled orbits. Application to
  magnesium}},\ }\href {https://doi.org/10.1103/PhysRev.147.505} {\bibfield
  {journal} {\bibinfo  {journal} {Phys. Rev.}\ }\textbf {\bibinfo {volume}
  {147}},\ \bibinfo {pages} {505} (\bibinfo {year} {1966})}\BibitemShut
  {NoStop}%
\bibitem [{\citenamefont {van Delft}\ \emph {et~al.}(2018)\citenamefont {van
  Delft}, \citenamefont {Pezzini}, \citenamefont {Khouri}, \citenamefont
  {Mueller}, \citenamefont {Breitkreiz}, \citenamefont {Schoop}, \citenamefont
  {Carrington}, \citenamefont {Hussey},\ and\ \citenamefont
  {Wiedmann}}]{VanDelft2018}%
  \BibitemOpen
  \bibfield  {author} {\bibinfo {author} {\bibfnamefont {M.~R.}\ \bibnamefont
  {van Delft}}, \bibinfo {author} {\bibfnamefont {S.}~\bibnamefont {Pezzini}},
  \bibinfo {author} {\bibfnamefont {T.}~\bibnamefont {Khouri}}, \bibinfo
  {author} {\bibfnamefont {C.~S.~A.}\ \bibnamefont {Mueller}}, \bibinfo
  {author} {\bibfnamefont {M.}~\bibnamefont {Breitkreiz}}, \bibinfo {author}
  {\bibfnamefont {L.~M.}\ \bibnamefont {Schoop}}, \bibinfo {author}
  {\bibfnamefont {A.}~\bibnamefont {Carrington}}, \bibinfo {author}
  {\bibfnamefont {N.~E.}\ \bibnamefont {Hussey}},\ and\ \bibinfo {author}
  {\bibfnamefont {S.}~\bibnamefont {Wiedmann}},\ }\bibfield  {title} {\bibinfo
  {title} {{Electron-hole tunneling revealed by quantum oscillations in the
  nodal-line semimetal HfSiS}},\ }\href
  {https://doi.org/10.1103/PhysRevLett.121.256602} {\bibfield  {journal}
  {\bibinfo  {journal} {Phys. Rev. Lett}\ }\textbf {\bibinfo {volume} {121}},\
  \bibinfo {pages} {256602} (\bibinfo {year} {2018})}\BibitemShut {NoStop}%
\bibitem [{\citenamefont {M{\"{u}}ller}\ \emph {et~al.}(2020)\citenamefont
  {M{\"{u}}ller}, \citenamefont {Khouri}, \citenamefont {van Delft},
  \citenamefont {Pezzini}, \citenamefont {Hsu}, \citenamefont {Ayres},
  \citenamefont {Breitkreiz}, \citenamefont {Schoop}, \citenamefont
  {Carrington}, \citenamefont {Hussey},\ and\ \citenamefont
  {Wiedmann}}]{Muller2020}%
  \BibitemOpen
  \bibfield  {author} {\bibinfo {author} {\bibfnamefont {C.~S.}\ \bibnamefont
  {M{\"{u}}ller}}, \bibinfo {author} {\bibfnamefont {T.}~\bibnamefont
  {Khouri}}, \bibinfo {author} {\bibfnamefont {M.~R.}\ \bibnamefont {van
  Delft}}, \bibinfo {author} {\bibfnamefont {S.}~\bibnamefont {Pezzini}},
  \bibinfo {author} {\bibfnamefont {Y.~T.}\ \bibnamefont {Hsu}}, \bibinfo
  {author} {\bibfnamefont {J.}~\bibnamefont {Ayres}}, \bibinfo {author}
  {\bibfnamefont {M.}~\bibnamefont {Breitkreiz}}, \bibinfo {author}
  {\bibfnamefont {L.~M.}\ \bibnamefont {Schoop}}, \bibinfo {author}
  {\bibfnamefont {A.}~\bibnamefont {Carrington}}, \bibinfo {author}
  {\bibfnamefont {N.~E.}\ \bibnamefont {Hussey}},\ and\ \bibinfo {author}
  {\bibfnamefont {S.}~\bibnamefont {Wiedmann}},\ }\bibfield  {title} {\bibinfo
  {title} {{Determination of the Fermi surface and field-induced quasi-particle
  tunneling around the Dirac nodal-loop in ZrSiS}},\ }\href
  {https://doi.org/10.1103/physrevresearch.2.023217} {\bibfield  {journal}
  {\bibinfo  {journal} {Phys. Rev. Res.}\ }\textbf {\bibinfo {volume} {2}},\
  \bibinfo {pages} {023217} (\bibinfo {year} {2020})}\BibitemShut {NoStop}%
\bibitem [{\citenamefont {M{\"{u}}ller}\ \emph {et~al.}(2022)\citenamefont
  {M{\"{u}}ller}, \citenamefont {{Van Delft}}, \citenamefont {Khouri},
  \citenamefont {Breitkreiz}, \citenamefont {Schoop}, \citenamefont
  {Carrington}, \citenamefont {Hussey},\ and\ \citenamefont
  {Wiedmann}}]{Muller2022}%
  \BibitemOpen
  \bibfield  {author} {\bibinfo {author} {\bibfnamefont {C.~S.}\ \bibnamefont
  {M{\"{u}}ller}}, \bibinfo {author} {\bibfnamefont {M.~R.}\ \bibnamefont {{Van
  Delft}}}, \bibinfo {author} {\bibfnamefont {T.}~\bibnamefont {Khouri}},
  \bibinfo {author} {\bibfnamefont {M.}~\bibnamefont {Breitkreiz}}, \bibinfo
  {author} {\bibfnamefont {L.~M.}\ \bibnamefont {Schoop}}, \bibinfo {author}
  {\bibfnamefont {A.}~\bibnamefont {Carrington}}, \bibinfo {author}
  {\bibfnamefont {N.~E.}\ \bibnamefont {Hussey}},\ and\ \bibinfo {author}
  {\bibfnamefont {S.}~\bibnamefont {Wiedmann}},\ }\bibfield  {title} {\bibinfo
  {title} {{Field-induced quasi-particle tunneling in the nodal-line semimetal
  HfSiS revealed by de Haas-van Alphen quantum oscillations}},\ }\href
  {https://doi.org/10.1103/PhysRevResearch.4.043008} {\bibfield  {journal}
  {\bibinfo  {journal} {Phys. Rev. Res.}\ }\textbf {\bibinfo {volume} {4}},\
  \bibinfo {pages} {043008} (\bibinfo {year} {2022})}\BibitemShut {NoStop}%
\bibitem [{\citenamefont {Breitkreiz}\ and\ \citenamefont
  {Brouwer}(2023)}]{Breitkreiz2022}%
  \BibitemOpen
  \bibfield  {author} {\bibinfo {author} {\bibfnamefont {M.}~\bibnamefont
  {Breitkreiz}}\ and\ \bibinfo {author} {\bibfnamefont {P.~W.}\ \bibnamefont
  {Brouwer}},\ }\bibfield  {title} {\bibinfo {title} {{Fermi-arc metals}},\
  }\href {https://doi.org/10.1103/PhysRevLett.130.196602} {\bibfield  {journal}
  {\bibinfo  {journal} {Phys. Rev. Lett}\ }\textbf {\bibinfo {volume} {130}},\
  \bibinfo {pages} {196602} (\bibinfo {year} {2023})}\BibitemShut {NoStop}%
\bibitem [{\citenamefont {Keller}(1958)}]{Keller1958}%
  \BibitemOpen
  \bibfield  {author} {\bibinfo {author} {\bibfnamefont {J.~B.}\ \bibnamefont
  {Keller}},\ }\bibfield  {title} {\bibinfo {title} {{Corrected Bohr-Sommerfeld
  Quantum Conditions for Nonseparable Systems}},\ }\href@noop {} {\bibfield
  {journal} {\bibinfo  {journal} {Ann. Phys. (N. Y).}\ }\textbf {\bibinfo
  {volume} {4}},\ \bibinfo {pages} {180} (\bibinfo {year} {1958})}\BibitemShut
  {NoStop}%
\bibitem [{\citenamefont {Landau}\ and\ \citenamefont
  {Lifshitz}(1977)}]{Landau1977}%
  \BibitemOpen
  \bibfield  {author} {\bibinfo {author} {\bibfnamefont {L.~D.}\ \bibnamefont
  {Landau}}\ and\ \bibinfo {author} {\bibfnamefont {E.~M.}\ \bibnamefont
  {Lifshitz}},\ }\href@noop {} {\emph {\bibinfo {title} {{Course of Theoretical
  Physics}}}}\ (\bibinfo  {publisher} {Elsevier},\ \bibinfo {address}
  {Oxford},\ \bibinfo {year} {1977})\BibitemShut {NoStop}%
\bibitem [{\citenamefont {Dwivedi}\ and\ \citenamefont
  {Chua}(2016)}]{Dwivedi2016b}%
  \BibitemOpen
  \bibfield  {author} {\bibinfo {author} {\bibfnamefont {V.}~\bibnamefont
  {Dwivedi}}\ and\ \bibinfo {author} {\bibfnamefont {V.}~\bibnamefont {Chua}},\
  }\bibfield  {title} {\bibinfo {title} {{Of bulk and boundaries: Generalized
  transfer matrices for tight-binding models}},\ }\href
  {https://doi.org/10.1103/PhysRevB.93.134304} {\bibfield  {journal} {\bibinfo
  {journal} {Phys. Rev. B}\ }\textbf {\bibinfo {volume} {93}},\ \bibinfo
  {pages} {134304} (\bibinfo {year} {2016})}\BibitemShut {NoStop}%
\bibitem [{\citenamefont {Dwivedi}\ and\ \citenamefont
  {Ramamurthy}(2016)}]{Dwivedi2016}%
  \BibitemOpen
  \bibfield  {author} {\bibinfo {author} {\bibfnamefont {V.}~\bibnamefont
  {Dwivedi}}\ and\ \bibinfo {author} {\bibfnamefont {S.~T.}\ \bibnamefont
  {Ramamurthy}},\ }\bibfield  {title} {\bibinfo {title} {{Connecting the dots:
  Time-reversal symmetric Weyl semimetals with tunable Fermi arcs}},\ }\href
  {https://doi.org/10.1103/PhysRevB.94.245143} {\bibfield  {journal} {\bibinfo
  {journal} {Phys. Rev. B}\ }\textbf {\bibinfo {volume} {94}},\ \bibinfo
  {pages} {245143} (\bibinfo {year} {2016})}\BibitemShut {NoStop}%
\bibitem [{\citenamefont {Groth}\ \emph {et~al.}(2014)\citenamefont {Groth},
  \citenamefont {Wimmer}, \citenamefont {Akhmerov},\ and\ \citenamefont
  {Waintal}}]{Groth2014}%
  \BibitemOpen
  \bibfield  {author} {\bibinfo {author} {\bibfnamefont {C.~W.}\ \bibnamefont
  {Groth}}, \bibinfo {author} {\bibfnamefont {M.}~\bibnamefont {Wimmer}},
  \bibinfo {author} {\bibfnamefont {A.~R.}\ \bibnamefont {Akhmerov}},\ and\
  \bibinfo {author} {\bibfnamefont {X.}~\bibnamefont {Waintal}},\ }\bibfield
  {title} {\bibinfo {title} {{Kwant: a software package for quantum
  transport}},\ }\href {https://doi.org/10.1088/1367-2630/16/6/063065}
  {\bibfield  {journal} {\bibinfo  {journal} {New J. Phys.}\ }\textbf {\bibinfo
  {volume} {16}},\ \bibinfo {pages} {063065} (\bibinfo {year}
  {2014})}\BibitemShut {NoStop}%
\end{thebibliography}%

\end{document}